\documentclass[aps,prd,twocolumn,nofootinbib]{revtex4-2}
\usepackage{graphicx} % Required to insert images
\usepackage{hyperref}
\usepackage{natbib}
\usepackage{ragged2e} % For left justification
\usepackage{amsmath}
\usepackage{color} % for the command \textcolor
\usepackage{soul} % for the command \hl
\begin{document}
%\title{Obtaining Nelson-Barr from P solution of the strong CP problem}
%\title{Electron Mass and Parity }
%\title{The One Where the Electron gets its Mass}

%\title{$P$ $\&$ $m_e$ }
\title{Parity and lepton masses in the left-right symmetric model}
%(Un)Likelihood of leptonic CP violation with Bayesian prior \\ motivated by the strong CP problem
\author{Ravi Kuchimanchi}
%\date{March 27, 2023}
\email{raviparity@gmail.com }

%\footnote{This paper has been published in Physical Review D with DoI https://doi.org/10.1103/PhysRevD.110.095028}

\begin{abstract}
%If parity (P) is restored in the laws of nature, then chiral ($\chi$) symmetry that sets the electron's mass to zero 
%(to explain its smallness) 
%also implies the vanishing of neutrino mixing angles in the minimal left-right symmetric model.  
Curiously in the minimal left right symmetric model, chiral symmetry that protects the electron's mass ($m_e$), due to parity (P), implies in the symmetry limit the vanishing of its neutrino mixing angles.  We break the chiral symmetry softly (or spontaneously if it is gauged) to generate the observed large neutrino mixing angles at the tree-level. The electron then acquires its mass on renormalization group equation (RGE) running due to its neutrino's mixing, and in turn determines the $B-L$ gauge symmetry breaking scale ($v_R$) to be $10^{10} GeV \lesssim v_R \leq 10^{15} GeV. $ If the muon's mass is also generated radiatively, the $B-L$ breaking scale is  $\sim 10^{14-15}$ GeV. Regardless of the high scale of $v_R$, this is a testable model since on RGE running and P breaking, a large strong CP phase ($\bar{\theta} >> 10^{-10}$) which depends logarithmically on $v_R$ is generated if there is $\mathcal{O}(1)$ CP violation in leptonic Yukawa couplings.  Hence we expect that leptonic CP phases including the Dirac CP phase $\delta_{CP}$ of the PMNS matrix must be consistent with $0$ or $180^o$ to within a degree, which can be verified or excluded by neutrino experiments such as DUNE and Hyper-Kamiokande. In lieu of P, if charge conjugation C is used, the same results follow. However with C and no P, axions would likely need to be added anyway, in which case there is no constraint on $\delta_{CP}$.  \\ \\
DOI: \href{https://doi.org/10.1103/PhysRevD.110.095028}{10.1103/PhysRevD.110.095028} \let\thefootnote\relax\footnote{Published in \href{https://doi.org/10.1103/PhysRevD.110.095028}{Phys. Rev. D \textbf{110}, 095028 (2024)}.}
%\addtocounter{footnote}{-1}\let\thefootnote\svthefootnote

\end{abstract}
\maketitle

\section{Introduction}
%\textit{Introduction.} 
\let\svthefootnote\thefootnote

It is well known that approximate chiral symmetries~\cite{FRITZSCH1978317, Altarelli_2000} that act on left handed first and second generation quarks, can suppress the associated quark masses and simultaneously also their Cabibbo-Kobayashi-Maskawa or CKM mixing (which is their left handed mixing), leading to an understanding of the smallness of both. 
However in the leptonic sector while the charged leptons exhibit mass hierarchy (notably, $m_e/m_\tau \sim 3 \times 10^{-4}$), the neutrino mixing angles in the Pontecorvo-Maki-Nakagawa-Sakata (PMNS) matrix (which is their left-handed mixing) are large or $\sim 0.1-1$  and are anarchical~\cite{PhysRevLett.84.2572, PhysRevD.63.053010} rather than hierarchical.

To accommodate neutrino masses and mixing, we can treat the Standard Model (SM) as an effective low energy theory and add a non renormalizable Weinberg term~\cite{PhysRevD.22.1694} $L_{iL}^T H H^T L_{jL}$, where $L_{iL}$ are leptonic $SU(2)_L$ doublets ($i = 1, 2, 3$ for the three generations) and $H$ is the SM Higgs doublet. Now the charged lepton mass hierarchy can be understood as being due to approximate chiral symmetry acting on the right handed charged leptons (rather than left handed), so as not to suppress the observed neutrino mixing~\cite{Altarelli_2004}.

If we augment the SM with an $SU(2)_L$ Higgs triplet $\Delta_L$~\cite{Bonilla:2015eha}, while the symmetry under $e_R \rightarrow e^{i\psi}e_R$ can be broken approximately to keep the electron mass ($m_e$) small, the observed large neutrino mixing can be generated by the Majorana type Yukawa terms $if_{ij} L_{iL}^T \tau_2\Delta_L L_{jL}$ when $\Delta_L$ picks up a VEV (Type II seesaw mechanism), with $f_{ij}$ %$\sim 0.1 - 1$ being unsuppressed 
being unaffected by the chiral symmetry. %$\sim v_{wk}^2/M$ where $v_{wk}$ is the weak scale and $M$ is its large mass.

There is however a vexing issue in the popular minimal Left-Right symmetric (LR) model~\cite{PhysRevD.10.275,*PhysRevD.11.566,*Senjanovic:1975rk,PhysRevD.44.837, Duka:1999uc},  %(with above triplet Higgs $\Delta_L$ and corresponding $\Delta_R$, and bi-doublet Higgs $\phi$ that contains two SM doublets) 
that restores parity as a good symmetry of nature. 

$e_R$ is now paired with $\nu_{eR}$ in the first generation $SU(2)_R$ lepton doublet $L_{1R}$, and parity (P or charge conjugation C) implies that if there is an approximate chiral symmetry acting on the right handed-leptons, there is a corresponding symmetry on the left-handed leptons. % $L_{1R} \rightarrow -L_{1R}$ (or  
The $U(1)(\equiv U(1)_R)$ symmetry,  $L_{1R} \rightarrow e^{i\psi} L_{1R}$  imposed to keep the electron massless, will due to parity, imply there is also a corresponding $U(1)_L$  $\chi-$symmetry  %$L_{1L} \rightarrow -L_{1L}$ % (or 
$L_{1L} \rightarrow e^{i\chi} L_{1L}$
which renders the above Majorana Yukawa couplings of $L_{1L}$ with $\Delta_L$, namely, $f_{1j} =f_{j1} =0$. % for $j=2,3$. 
To generate unsuppressed $f_{1j} = f_{j1}$ the chiral symmetry must be broken in a non-approximate way. % Note that we can also use the chiral symmetry $L_{1L} \rightarrow e^{i\chi} L_{1L}$ in place of the discrete $\chi$ symmetry that also makes $f_{11}=0$ and is more predictive. 
%In other words, 't Hooft's test for naturalness for electron mass would require setting  to zero 
%not only the electron mass but also its neutrino's mixing 
%not only the Dirac type but also Majorana type Yukawa couplings of the first generation. The symmetry that is then restored is the $\chi-$ symmetry, which motivates us to explore it further.

As far as we know, this crucial aspect of
%central issue surrounding Type II dominant seesaw mechanism and the electron mass 
 the minimal left right symmetric model hasn't been adequately discussed and addressed before. %and this is the first effort that has Type II dominant seesaw. 
 This work is the first to explore it within the framework of the Type II dominant seesaw mechanism.  %Alternately, if $f_{1j}$ are suppressed, because of  this issue due to P,  we would need correlated cancellations between the numerator and  denominator of Type I seesaw mechanism which is not very satisfactory.  

That is, in the minimal left right symmetric model if we make the electron massless by restoring a chiral symmetry,
%the above Majorana Yukawa couplings $f_{12}, f_{13}$ of the first generation with $\Delta_L$ also vanish in the symmetry limit.  And thus 
its neutrino mixing also vanishes, and it is a challenge to understand the observed largeness of the leptonic mixing angles~(see also \cite{Altarelli_2004}, and 't Hooft's naturalness criterion - setting a small parameter, or set of parameters, such as electron mass to zero should restore a symmetry~\cite{Hooft1980}).

In this work we address the challenge  by adding a second triplet $\Delta'_L$ ($\rightarrow e^{-i\chi}\Delta'_L$ under $U(1)_L$  so that it can couple to $L_{1L}$), whose mixing with $\Delta_L$ breaks the $\chi-$symmetry and generates the observed neutrino mixing angles of the first generation, while the electron remains massless at the tree level.  After integrating out the heavy $\Delta'_L$  (by going to its mass basis at its mass scale $M'$) and its parity partner $\Delta'_R$,  we obtain the minimal left right symmetric model which on RGE running to the parity breaking scale $v_R < M'$,  generates the electron mass (corresponding Yukawa coupling) in two loops, and determines $v_R$.  Thus we not only gain an understanding of the smallness but also obtain the electron's mass from its neutrino mixing and a range for the scale of $v_R$.

%Since we already know the electron's mass we can turn the argument around into a prediction that the P or $SU(2)_R \times U(1)_{B-L}$ breaking scale is $v_R\sim 10^{11-15}$ GeV. 

The idea of generating the masses and mixing angles of the first and/or second generation leptons and/or quarks radiatively at different loop orders has been previously explored in different contexts over the years~\cite{Balakrishna:1987qd, Balakrishna:1988ks, Balakrishna:1988xg,  Barr:1990td, Dobrescu:2008sz, Babu:2020bgz, PhysRevD.106.075020}.  In some of the past work, the scalar content has been chosen so that there are no gauge invariant Yukawa terms that directly give masses to lighter fermions.  %The scalar content is usually chosen so that gauge symmetry prevents Yukawa couplings that can give tree level masses to fermions.  
%Several heavy fields that interact with the light fermions are added to enable their radiative mass generation.
In the minimal Left Right symmetric model, as also the SM,  the scalar content is such that Yukawa terms that give charged fermion masses are allowed by gauge symmetry.  In this work we impose chiral symmetry (which can also be gauged) to protect the electron mass, which gets generated on RGE running from neutrino mass and mixing parameters of the minimal left-right symmetric model itself, below the  chiral symmetry breaking scale.

Our model is testable, since as shown in Section~\ref{sec:test}, in the absence of axions,   $O(1)$ CP violating phases in leptonic Yukawa couplings generate too large a strong CP phase in one loop, and therefore the leptonic CP phase $\delta_{CP}$ of the PMNS matrix which will be measured by neutrino experiments DUNE and Hyper-K must be absent (negligibly small) mod $\pi$. 

The rest of the paper is organized as follows. In Section~\ref{sec:chi} we impose chiral symmetry to set the electron's Yukawa couplings to zero, and break the symmetry softly (or spontaneously as in Section~\ref{sec:gauge}) so as to generate the electron neutrino's mixing at the tree level.  In Section~\ref{sec:emass} we show that the neutrino mixing causes the electron's mass to be radiatively generated in two loop RGE running, which in turn determines the parity or $B-L$ breaking scale $10^{10} GeV \lesssim v_R \lesssim 10^{15} GeV$. Section~\ref{sec:mu} discusses how the mechanism can be extended to also generate the muon mass radiatively. Section~\ref{sec:test} shows how our model with $P$ can be tested by upcoming neutrino experiments. Section~\ref{sec:C} discusses the case where there is $C$ rather than $P$.   Section~\ref{sec:gauge} shows that the chiral $U(1)_L \times U(1)_R$ symmetry that we imposed in Section~\ref{sec:chi} can be gauged. And we summarize the conclusions in Section~\ref{sec:conc}.   
\section{Chiral symmetry breaking \& neutrino mixing}
\label{sec:chi}
%\textit{Chiral symmetry breaking \& neutrino mixing.} 
We begin with the Left-Right symmetric model~\cite{PhysRevD.10.275,*PhysRevD.11.566,*Senjanovic:1975rk,PhysRevD.44.837, Duka:1999uc} based on $SU(3)_c \times SU(2)_L \times SU(2)_R \times U(1)_{B-L} \times P$ with the usual Higgs triplet $\Delta_R $, augmented by an additional $\Delta'_R$ and parity partners $\Delta_L, \Delta'_L$, along with bi-doublet $\phi $.
%(for the Higgs potential please see for example~\cite{Duka:1999uc,ZHANG2008247}).    

 Scalars and leptons in our model are displayed in Table~\ref{tab:matterta} and there are also the quarks $Q_{iL}$ (doublet of $SU(2)_L$) and $Q_{iR}$ ($SU(2)_R$ doublet) with $B-L = 1/3$.  Note that the left handed electron $e_L \equiv e^-_{1L}$ and its neutrino $\nu_{eL} \equiv \nu_{1L}$ are in the first generation doublet $L_{1L}$. And likewise the left-handed muon and tau generations are $L_{2L}$ and $L_{3L}$ respectively. 

%The matter content is displayed in table~\ref{tab:matterta} where the SM right handed $SU(2)_L$ singlet fermions, are all in the corresponding $SU(2)_R$ doublets $Q_{iR}$ and $L_{iR}$,  the right-handed neutrinos are in $L_{iR}$, and the subscript $i = 1, 2, 3$ represents the 3 generations. 

Under Parity ($P$), spacetime $(x,t) \rightarrow (-x,t),$ the gauge bosons of $SU(2)_L \leftrightarrow SU(2)_R$,   $\phi \rightarrow \phi^\dagger$, $Q_{iL} \leftrightarrow Q_{iR}$ for the quarks,  and similarly subscripts $L \leftrightarrow R$ for all the  fields in table~\ref{tab:matterta}. $P$ is broken spontaneously at scale $v_R$.

Note that instead of $P$ we can impose $C$ which also exchanges $SU(2)_L$ and $SU(2)_R$ groups.  Matrices $h^\ell$ and $\tilde{h}^\ell$ in equation~(\ref{eq:hyuk}) will then be symmetric instead of being Hermitian, and lead to essentially the same calculations and results for the electron mass and $v_R$ scale. We will proceed using $P$, and please also read Section~\ref{sec:C} for $C$.

\begin{table}[t]
\centering
\begin{tabular}{lc}
\hline 
& 
$\scriptstyle{SU(2)_L\times SU(2)_R\times U(1)_{B-L}}$
\\  \hline \\
$\Delta_L = \left(
\begin{array}{cc}
\Delta_L^+/\sqrt{2} & \Delta_L^{++} \\
\Delta_L^o  & -\Delta^+_L/\sqrt{2}
\end{array}
\right), \ \Delta'_L$ & $(3,1,2)$ \\ 
$\Delta_R = \left(
\begin{array}{cc}
\Delta_R^+/\sqrt{2} & \Delta_R^{++} \\
\Delta_R^o  & -\Delta^+_R/\sqrt{2}
\end{array}
\right), \ \Delta'_R  $ & $(1,3,2)$ \\ 
$\phi = \left(
\begin{array}{cc}
\phi^o_1 & \phi_2^+ \\
\phi^-_1  & \phi^o_2
\end{array}
\right)$  & $(2,2,0)$ \\ %\\ \hline \\
%$Q_{iL} $ & $(3,2,1,1/3)$ \\ 
%$Q_{iR} $ & $(3,1,2,1/3)$ \\ 
$L_{iL} = \left(
\begin{array}{c}
\nu_{iL}  \\
e^{-}_{iL} 
\end{array}
\right) $, $L_{iR} =\left(
\begin{array}{c}
\nu_{iR}  \\
e^{-}_{iR} 
\end{array}
\right) $ & $(2,1,-1), (1,2,-1)$ \\ 
%$Q_{iL} = \left(
%\begin{array}{c}
%u_{iL}  \\
%d_{iL} 
%\end{array}
%\right) $ & $(2,1,-1)$ \\ 
%$Q_{iR} =\left(
%\begin{array}{c}
%u_{iR}  \\
%d_{iR} 
%\end{array}
%\right) $ & $(1,2,1/3)$ 
\\ \hline 
\end{tabular}
\caption{Scalars and leptons of the minimal left right symmetric model with additional scalar triplets $\Delta'_{L,R}$ (whose $2\times 2$ matrix representation is similar to $\Delta_{L,R}$). Neutral components of scalars with superscript ``$o$"  pick up VEVs. }
\label{tab:matterta}
\end{table}

We impose chiral symmetry $\chi$ under which 
\begin{equation}
L_{1L} \rightarrow e^{i\chi}L_{1L}, \ \Delta'_L \rightarrow e^{-i\chi}\Delta'_L. 
\label{eq:chi}
\end{equation}
This is an $U(1)_L$  symmetry, and due to P we automatically have corresponding  $U(1)_R$, or  $U(1)_L \times U(1)_R$ global  symmetry (we later show in Section~\ref{sec:gauge}  that this can be gauged).   Alternatively the chiral symmetry $\chi$  can be the discrete
\begin{equation}
L_{1L} \rightarrow -L_{1L}, \ \Delta'_L \rightarrow -\Delta'_L.
\label{eq:chidis}
\end{equation}  
%The SM group is broken by the VEVs $\left<\phi^0_1\right> \equiv \kappa_1$ and $\left<\phi^0_2\right> \equiv \kappa_2$  of the bidoublet $\phi$ (with the weak scale $v_{wk}^2 = |\kappa_1|^2 +|\kappa_2|^2$).  $\phi$ has two SM Higgs doublets labeled below by subscripts 1 and 2,  and can be represented by the matrix 

The Yukawa couplings that give Dirac type mass terms to the leptons have the usual form
\begin{equation}
    h^\ell_{ij} \bar{L}_{iL} \phi L_{jR} + \tilde{h}^\ell_{ij} \bar{L}_{iL} \tilde{\phi} L_{jR} + h.c.
    \label{eq:hyuk}
\end{equation}
where $\tilde{\phi} = \tau_2 \phi^\star \tau_2 = \left(
\begin{array}{cc}
\phi^{o\star}_2 & -\phi_1^+ \\
-\phi^-_2  & \phi^{o\star}_1
\end{array}
\right),$ and $h^\mathnormal{\ell}_{ij} = h_{ji}^{\ell\star}$,  $\tilde{h}^\ell_{ij} = \tilde{h}_{ji}^{\ell\star}$ are elements of Yukawa matrices $h^\ell$ and $\tilde{h}^\ell$, where the Hermiticity is due to P, and $\tau_2$ is the Pauli matrix. Note that the superscript $\ell$ denotes that these are leptonic Yukawa couplings (of the Dirac type).  

The chiral symmetry $\chi$ (with P) implies for all $j$ that
\begin{equation}
    h^\ell_{1j} = h^\ell_{j1} = \tilde{h}^\ell_{1j} = \tilde{h}^\ell_{j1} = 0.
\label{eq:hij}
\end{equation}
Which implies that the electron's Yukawa couplings that generate its mass vanish in the symmetry limit (that is, above the chiral symmetry breaking scale).

The Majorana type Yukawa terms are of the form (note that $L \rightarrow R$ refers to the subscript $L$),
\begin{equation}
    if_{ij} {L}_{iL}^T \tau_2 \Delta_L L_{jL} +  if'_{ij} {L}_{iL}^T \tau_2 \Delta'_L L_{jL} + L\rightarrow R + h.c. 
    \label{eq:majyuk}
\end{equation}
where $f, f'$ are symmetric matrices ($f_{ij} = f_{ji}, f'_{ij} = f'_{ji}$), and due to $\chi$ symmetry, 
\begin{equation}
    f_{12} = f_{13} = f'_{33} = f'_{22} = f'_{11} = f'_{23} = 0.
    \label{eq:f=0}
\end{equation}
$f_{11}= 0$ (or $f_{11}\neq 0)$) depending on eq.~(\ref{eq:chi}) (or eq.~(\ref{eq:chidis})).

%\begin{eqnarray}
%V = & M^2 & Tr (\Delta_L^\dagger \Delta_L)  + M'^2 Tr (\Delta'_L^\dagger \Delta'_L) +   \nonumber \\ & + & Tr (\mu'^2 \Delta'_L^\dagger \Delta_L + h.c.) + L \rightarrow R
%\end{eqnarray}
%with $\mu'^2, M'^2, M^2 \geq v_R$.

%Writing the VEVs of neutral components of $\left<\Delta^o_R\right>\sim v_R$ and the bidoublet $\left<\phi^o\right> \sim v_{wk}$, we note that the $\beta$ term in the Higgs potential $\beta Tr(\Delta_R^\dagger \phi \Delta_L \phi)$ induces a VEV $v_L \sim \beta v_{wk}^2/v_R$ to $\Delta^o_L$ that in turn induces a VEV to $\Delta'^o_L \sim (\mu'^2/M'^2) v_L \sim (0.1-1)v_L$, where the superscript $o$ denotes the neutral components. 

%Thus the neutrino mass 

The mass terms  of the Higgs potential that involve the scalar triplets can be compactly written as 
\begin{equation}
V_{mass} =   Tr \left[(\Delta_L^\dagger \ \Delta'^\dagger_L) \mathcal{M}^2 \left(
\begin{array}{c}
    \Delta_L \\
\Delta'_L
\end{array}
\right)\right] + L \rightarrow R
\label{eq:vmass}
\end{equation}
%\begin{equation}
with $\mathcal{M}^2=
\left(
\begin{array}{cc}
    M^2 & \mu'^{2\star} \\
\mu'^2 & M'^2
\end{array}
\right)
\label{eq:m^2}$
%\end{equation}
where $\mu'^2$ breaks the $\chi$ symmetry softly (since it is a mass dimension 2 term), and $\mathcal{M}^2$ is Hermitian. Hereafter we  let $\mu'^2$  be real as its phase can be absorbed into a redefinition of the $\Delta'_{L,R}$ fields. 

Note that $\mu'^2$ can be generated from spontaneous breaking of discrete or gauged $\chi$ symmetry as shown in Sec.~\ref{sec:gauge}. 

We take $\mu'^2 /M'^2 \sim (0.1~to~0.8)$ (require $\mathcal{O}(1)$, or large-enough, $\chi$ symmetry breaking to generate sufficient $\nu_e$ mixing, and radiatively the electron mass) and we choose $\Delta'_L$ (and therefore also $\Delta'_R$) to have a mass $\sim M'^2 >> v_R^2$ so that it decouples and we just have the minimal left right symmetric model with $P$ below the scale $M'$. That is,  $\mathcal{M}^2$ has two eigenvalues,  the heavier being $\sim Tr \mathcal{M}^2\sim M'^2 + M^2 \sim M'^2$ (we take $M^2 \leq M'^2$ without loss of generality) and the lighter that we rename as $-\mu_3^2 \sim -v_R^2$. The lighter eigenvalue 
$-\mu_3^2$ is negative as it generates the VEV $v_R < M'$. % which breaks $SU(2)_R \times U(1)_{B-L}$.

Evaluating $Det\mathcal{M}^2 \sim (-v_R^2)(M'^2)$ using the above form of $\mathcal{M}^2$,  we obtain $M^2 - |\mu'|^4/M'^2 \sim -v_R^2$.  This is the usual fine-tuning due to the hierarchy problem associated with keeping the $SU(2)_R \times U(1)_{B-L}$ gauge symmetry breaking scale $v_R$  smaller than the higher mass scales or cut-off scale\footnote{If we supersymmetrize~\cite{PhysRevD.48.4352}, then this fine-tuning issue disappears which also indicates that it is the usual gauge hierarchy problem.  For example the superpotential $W = (\mu' \Delta + M' \Delta') \overline{\Delta}'$ that is symmetric under $\overline{\Delta} \rightarrow -\overline{\Delta}$ mixes the superfields $\Delta$ and $\Delta'$ at scale $M'$ while the mass of $\Delta, \overline{\Delta}$ (renaming the lighter mass eigenstates to be these) is zero due to this symmetry and is well separated from $M'$. Note that $\mu'$ softly breaks the $\chi$ symmetry $L_{1}, \overline{\Delta}' \rightarrow  e^{i\chi}\{L_{1}, \overline{\Delta}'\}$, \ $\Delta' \rightarrow e^{-i\chi} \Delta'$ at scale $\sim M'$. % If we add a subscript L to the superfields they can be seen as the corresponding fields of our non-SUSY model.
}. There is no fine tuning we do other than what is anyway necessary for the hierarchy of the gauge symmetry breaking scales. %Typically $v_R \sim 10^{15} GeV$ which is the canonical seesaw scale, and $M'  \sim 10^{16~to~18} GeV$.

We now obtain the $2 \times 2$ orthogonal transformation $\mathcal{O}$ that diagonalizes $\mathcal{M}^2$ so that  $\mathcal{M}^2_{diag} = \mathcal{O}^T \mathcal{M}^2 \mathcal{O}$.  Note that the $\mathcal{O}_{21}$ term $sin \theta_\Delta$ depends on $\mu'^2 /M'^2$. We now change variables in the original Lagrangian  %Using $\mathcal{O}$ we rotate in the $\Delta_L$, $\Delta'_L$ plane (and likewise for $\Delta_R, \Delta'_R$) to go to the physical mass basis so that
\begin{equation}
\left.
\begin{array}{ll}
     \Delta_{L}\rightarrow cos(\theta_\Delta) \Delta_L - sin(\theta_\Delta) \Delta'_L \\
     \Delta'_L \rightarrow  sin(\theta_\Delta) \Delta_L + cos (\theta_\Delta) \Delta'_L  
    \end{array}
    \right\}  \ \ \& \ \ L \rightarrow R
    \label{eq:massbasis}
\end{equation}
where the fields on the right hand side are in the physical mass basis.

%Substituting for $\Delta_{L,R}, \Delta'_{L,R}$ in equation~(\ref{eq:majyuk}) the right hand side of the transformation in~(\ref{eq:massbasis}),
%\begin{equation}
%f_{ij} \rightarrow f_{ij} cos(\theta_\Delta) -  f'_{ij} sin(\theta_\Delta)
%\end{equation}
%and after that dropping fields $\Delta'_{L}, \Delta'_R$ as they have decoupled (and doing the same for all the Higgs potential terms), we find that below the mass given by the heavier eigenvalue $\sim M'$, we just have the lighter triplets $\Delta_{L,R}$  whose Yukawa couplings we can now rename to be $f_{ij}$.  

That is after substituting~(\ref{eq:massbasis}) into~(\ref{eq:majyuk}) and then dropping terms containing $\Delta'_{L,R}$ as they have decoupled (and doing the same for all the Higgs potential terms as well),  we rename the resulting Majorana Yukawa couplings of $\Delta_{L,R}$ as follows:
\begin{equation}
 \left[f_{ij} cos(\theta_\Delta) +  f'_{ij} sin(\theta_\Delta)\right] \rightarrow f_{ij}.
\end{equation}
 $f_{12}, f_{13}$ which were zero (see equation~(\ref{eq:f=0})) are now unsuppressed, and equation~(\ref{eq:majyuk}) becomes
\begin{equation}
    if_{ij} {L}_{iL}^T \tau_2 \Delta_L L_{jL}  + L\rightarrow R + h.c. 
    \label{eq:majyukminimalLR}
\end{equation}
with all $f_{ij}$ unsuppressed (except for $f_{11}=0$ in case of eq.~(\ref{eq:chi})), and $\Delta'_{L,R}$ have decoupled. % The mass term of the surviving $\Delta_{L,R}$ is hereafter written in the standard notation used in the Higgs potential of the left right symmetric model as $-\mu_3^2 (Tr \Delta_R \Delta_R^\dagger + R \rightarrow L)$, and the VEV $\left<\Delta^o_R\right> \sim v_R \sim \mu_3$  breaks $SU(2)_R \times U(1)_{B-L}$ to $U(1)_Y$. 

Importantly, equation~(\ref{eq:hij}) now only holds as a boundary condition at the $\chi$ symmetry breaking scale $M'$, and below that scale $h^\ell_{1j}, \tilde{h}^\ell_{1j}$ (responsible for the electron's mass) are generated from $h^\ell_{33}, \tilde{h}^\ell_{33}$ by renormalization group running (between scales $M'$ and $v_R$) due to the now non-zero $f_{12}$ and $f_{13}$. %Thus there is the exciting possibility of generating the electron mass from the $\tau$ Yukawa couplings and neutrino mixing parameters.

Note that $f_{11}=0$ [in the case of Eq.~\ref{eq:chi}] points to a normal mass hierarchy with the lightest neutrino getting its mass $\sim \sqrt{\Delta m_{12}^2}  sin^2\theta_{12} \approx 0.003 eV$ via the PMNS mixing angle from its heavier counterpart. %If in lieu of the continuous chiral symmetry in equation~(\ref{eq:chi}) we had a discrete $L_{1L}, \Delta'_L \rightarrow -L_{1L}, -\Delta'_L$, then $f_{11} \neq 0$.

Once $\Delta^o_R$, $\phi_1^o$ and $\phi_2^o$ (see Table~\ref{tab:matterta}) pick up VEVs $v_R, \kappa_1$ (can always be chose real), and $\kappa_2$ (taken real for ease of calculation, and we also note that its tree-level value is real for the axionless solution to the strong CP problem) respectively, the neutral component $\Delta^o_L$  picks up an induced VEV~\cite{PhysRevD.44.837} 
\begin{equation}
v_{L} = \left<\Delta^o_L \right> \sim (\beta_1\kappa_1\kappa_2+\beta_2 \kappa_1^2 + \beta_3 \kappa_2^2)/(\rho_3-2\rho_1)v_R
\label{eq:vL}
\end{equation}
because of the following terms in the Higgs potential 
\begin{equation}
\beta_1 Tr(\phi \Delta_R \phi^\dagger \Delta_L^\dagger)+\beta_2 Tr(\tilde{\phi} \Delta_R \phi^\dagger \Delta_L^\dagger)+\beta_3 Tr(\phi \Delta_R \tilde{\phi}^\dagger \Delta_L^\dagger)+ hc
\label{eq:beta2}
\end{equation}
where the $\beta_i$ are real due to Parity and $(\rho_3-2\rho_1) v^2_R$ is the mass of $\Delta_L$, with $\rho_3, \rho_1, \beta_i$ and $\mu^2_3$ (mentioned earlier) being the usual Higgs potential parameters of the minimal left-right symmetric model using the standard notation given for example in~\cite{PhysRevD.44.837, Duka:1999uc, AKHMEDOV2024138616}.

The charged lepton mass matrix $(h^{-})(v_{wk})$ and the light neutrino mass matrix $m_\nu$ can be evaluated using the Yukawa couplings in equations~(\ref{eq:hyuk}),(\ref{eq:majyukminimalLR})~\cite{PhysRevD.44.837}: 
%\begin{equation}
%\begin{center}
\begin{eqnarray}
%\begin{align}
h^{-} v_{wk} =  \ h^\ell \kappa_2 + \tilde{h}^\ell \kappa_1, \ \ \ \
 h^D v_{wk} =  \  h^\ell \kappa_1 + \tilde{h}^\ell \kappa^\star_2 \ \ \
%\end{aligned}
\label{eq:diracyuk} \\
%\end{equation}
%\begin{equation}
     m_\nu =  \ f v_L - \left(v^2_{wk}/v_R \right) h^D \left(\frac{1}{f} \right) h^{D^T} \ \ \ \ \
     \label{eq:mnu}
%     \end{align}
  \end{eqnarray}
 %    \end{center}
%\end{equation}
with equation~(\ref{eq:hij}) being the boundary condition at the scale $M'$, and where the weak interaction scale $v_{wk}^2 = \kappa_1^2 + |\kappa_2|^2 \sim \kappa^2_1$ (since the top quark gets its mass from $\kappa_1$), $v_R > \kappa_1 > v_L$, and $v_L$ is from equation~(\ref{eq:vL}). 

Note that the second term on the right hand side of equation~(\ref{eq:mnu}) for $m_\nu$ is the Type I seesaw contribution.  Substituting from equation~(\ref{eq:hij}) in $h^D$ of eqn~(\ref{eq:diracyuk}), we can see that the Type I seesaw term in eqn.~(\ref{eq:mnu}) does not contribute to the first generation's PMNS matrix elements at the scale $M'$ and therefore we have  Type II (first term of eq.~(\ref{eq:mnu})) dominant seesaw mechanism.  While it can contribute, for example to the third generation neutrino mass if $h^D_{33}$ is competitively large, it doesn't change our results, and therefore for the ease of calculations, without loss in generality, we ignore any Type I contribution from $h^D$.  

%The electron mass is generated by RGE running of $h^{-}$ from $M'$ scale to $v_R$ scale.  Since we will only be doing an estimate and the terms in the RGE that generate its mass are given by the same function that is linear in $h^\ell$ and $\tilde{h}^\ell$, the RGE running of $h$  

%\section{One loop RGE - No mass generation}
%\label{sec:oneloop}
\section{RGE, electron mass \& $v_R$ scale. } 
\label{sec:emass}
%\textit{RGE, electron mass \& $v_R$ scale. } 
The charged lepton Yukawa matrix $h^-$ (that determines the electron's mass) given in equation~(\ref{eq:diracyuk}) can be evaluated at the scale $v_R$  by using the RGEs of $h^\ell$ and $\tilde{h}^\ell$ which can be written in terms of their beta functions as
\begin{equation}
    \frac{\partial h^\ell}{\partial t} = \beta_{h^\ell} = \beta^{(1)}_{h^\ell} + \beta^{(2)}_{h^\ell} + ...
   \label{eq:betasums}
   \end{equation}
   and similarly the beta function $\beta_{\tilde{h}^\ell}$ with $h^\ell \rightarrow \tilde{h}^\ell$.   $t = ln \mu$ is the renormalization scale (and $\mu$ runs from the higher scale $M'$ down to $v_R$)  and the superscripts $(1)$ etc stand for the loop expansion.

   The one loop beta functions of the minimal left right symmetric model~\cite{CHAKRABORTTY2016361} include the term $\beta^{(1)}_{h^{\ell}} = (3/16 \pi^2) (f^\dagger f h^\ell + h^\ell f^\dagger f), %\label{eq:beta}
   $ and therefore $h^\ell_{13}=h^{\ell\star}_{31}, h^\ell_{12}=h^{\ell\star}_{21}$ which were $0$ at scale $M'$, get generated on RGE running from components such as $f_{13}^\star f_{33} h^\ell_{33}, h^\ell_{33}f_{33} ^\star f_{31}$.  This suggests that the Yukawa matrix $h^-$ constrained at scale $M'$ by equation~(\ref{eq:hij}) to have a zero eigenvalue (corresponding to massless electron) may pick up a small non-zero one loop contribution to its determinant on RGE running.

However on integrating the RGE equations numerically using the program $R$, we find  this not to be the case. Note that this one loop term is due to the correction in the propagator of the  fermions rather than being a true vertex correction.

%and the electron does not get a mass at the one-loop level.  
%Moreover if we set both the electron and the muon's Yukawa couplings to zero so that only $h^\ell_{33}$ and $\tilde{h}^\ell_{33}$ in equation~(\ref{eq:hyuk}) are non-zero, 
%though it may appear from the beta function in equation~(\ref{eq:beta}) and substituting from~(\ref{eq:majyukminimalLR}) that $h^\ell_{23}=h_{32}^{\ell\star}, h^\ell_{13}=h^{\ell\star}_{31}$ will be generated thereby making one of the zero eigenvalues of $h^\ell$ nonzero, yet 
%on RGE running we find that both the electron and muon remain massless to the one loop order.

%Thus we need to go to two loops to generate the Yukawa term that gives charged fermion masses.

%\section{Electron mass \& Seesaw scale}

\begin{figure}
    \centering
    \includegraphics[width=\columnwidth]{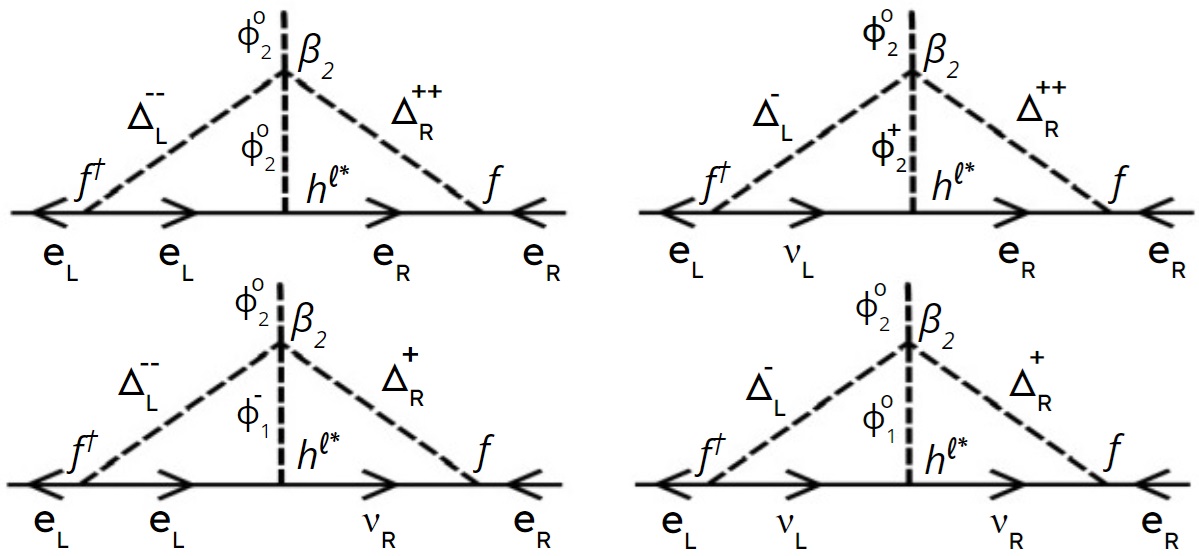} % Adjust width as needed
    \caption{Two loop contribution that generates $m_e$ from $\beta_2$ and $h^\ell$ (via eqs.~(\ref{eq:betatwoloop}),(\ref{eq:me})) when $\phi^o_2$ picks the VEV $\kappa_2$. % (that is from $h^-$ Yukawa coupling in eqn~(\ref{eq:diracyuk})).  
    The contribution from $\beta_1$ and $h^\ell$ to $\beta^{(2)}_{\tilde{h}^\ell}$ can be obtained from similar diagrams on relabelling $\beta_2 \rightarrow \beta_1$ and flipping the external scalar line from $\phi^o_2$ to $\phi^o_1$ which picks the VEV $\kappa_1$ (eq.~(\ref{eq:beta1mass})). %so that $h^\ell$ contributes to  and the electron mass via the VEV $\kappa_1$. %(that is from $h^D$ term in  eqn~(\ref{eq:diracyuk})).  Contributions from $\tilde{h}$ to $h^-$ and $h^D$ can be similarly written.
    }
    \label{fig:beta2}
\end{figure}

$h^\ell$ receives  two-loop contributions from the diagrams in Fig~\ref{fig:beta2} corresponding to the beta function term
\begin{equation}
  \beta^{(2)}_{h^{\ell}}   \sim {{6\beta_2} \over {(16 \pi^2)^2}}  \left(f^\dagger h^{\ell^\star} f \right)
    \label{eq:betatwoloop}
\end{equation}
%and is linear in $h^\ell$ (and likewise for $\tilde{h}^\ell$). The above term is the same function for both the Yukawa matrices.  Therefore $h^-$ in equation~(\ref{eq:diracyuk}) also has the same term governing its RG evolution, and relative values of $\tilde{h}^\ell\kappa_1$ and $h^\ell\kappa_2$  do not matter. That is, we can replace $h^\ell$ by $h^-$ in equation~(\ref{eq:betatwoloop}) to obtain its RGE, and from eigenvalues of $h^-$ at scale $v_R$, we obtain for the mass of the electron:  
and on RGE running contributes to $h^-$ in eqn~(\ref{eq:diracyuk}), and generates the mass of the electron
%\begin{figure}
%    \centering
%    %\includegraphics[width=\columnwidth]{feyn-final-use.jpg} % Adjust width as needed
%    \includegraphics{feyn-final-use.jpg}
%    \caption{Two loop contribution that generates the electron mass from $\beta_2$ and $h^\ell$ when $\phi^o_2$ picks the VEV $\kappa_2$. % (that is from $h^-$ Yukawa coupling in eqn~(\ref{eq:diracyuk})).  
%    The contribution from $\beta_1$ can be seen from above figures by relabelling $\beta_2 \rightarrow \beta_1$ and flipping the external scalar line from $\phi^o_2$ to $\phi^o_1$ so that $h^\ell$ contributes to the electron mass via the VEV $\kappa_1$. %(that is from $h^D$ term in  eqn~(\ref{eq:diracyuk})).  Contributions from $\tilde{h}$ to $h^-$ and $h^D$ can be similarly written.
 %   }
  %  \label{fig:beta2}
%\end{figure}

\begin{equation}
 m_e \approx \frac{6  h^\ell_{33} \kappa_2}{(16 \pi^2)^2} c_{\nu mix} \left( f_{33}^2 \beta_2\right) ln(M'/v_R)%, \ \ m_\tau \approx h^\ell_{33} \kappa_2  
 \label{eq:me}
\end{equation}
where we work in a basis where at scale $M'$,  $h^-$ is diagonal with $h^-_{1j} = h^-_{j1} = 0$ due to equation~(\ref{eq:hij}), to simplify calculation we assume $h^D$ is diagonal and also neglect Type I contribution,  and the matrix $f$ in equation~(\ref{eq:betatwoloop}) (and~\ref{eq:majyukminimalLR})) is almost fully determined by the first term on the right hand side of eq~(\ref{eq:mnu}) (up to a scale factor which we take to be $f_{33}$) using the known neutrino mass squared differences and PMNS mixing angles ($m_\nu$ determines $\hat{f}$ where $f=f_{33}\hat{f}$, and  $\hat{f}_{33}=1$).
%and with $f_{33} = 1$ to set the scale 
We have assumed normal ordering and the smallest neutrino mass is negligible.  %Note that we have used $m_\nu \approx f v_L$.% and $f_{33}$ can be used as the representative scale parameter. 

$c_{\nu mix} \approx c sin\theta_{13} \approx 0.1$ is the factor obtained on solving the RGE equations numerically using R, and is due to the PMNS mixing.  Keeping other things the same, if we half or double the PMNS mixing angle $sin\theta_{13}$ the above contribution to $m_e$ halves or doubles.  

We obtain $c \approx 1$ due to the observed values of the larger two mixing angles.  Had these larger PMNS mixing angles been zero, so would $c$ be, and $c_{\nu mix}$ would have scaled as $sin^2\theta_{13}$. It is interesting that for sufficient electron mass to be radiatively generated, the PMNS mixing needs to be reasonably large, as is the case in nature.

Substituting $c_{\nu mix} \approx 0.1$ and $h^\ell_{33}\kappa_2 \approx m_\tau \approx 1.8 GeV$ (it cant be much higher as it contributes to $\tau$ mass, and lower values produce lesser $m_e$) 
we obtain from eq.~(\ref{eq:me})

\begin{equation}
 m_e = 0.5 MeV \left(f_{33}^2 \beta_2\right) \left[\frac{ln(M'/v_R)}{10}\right]
 \label{eq:masselec}
 \end{equation}

 Taking $ln(M'/v_R)$ to be $\sim 1$ to $10$, the observed electron mass of 0.5 MeV is generated if 
 \begin{equation}
      f_{33}^2 \beta_2 \approx 1~~(to~10)
      \label{eq:one}
 \end{equation}
%This is also consistent with the fact that we naturally expected these couplings to be $\sim 1$ as we haven't imposed any symmetries that restrain them.
Since we require $f_{33}$ and $\beta_2$ to be within perturbation limits, neither can be much larger than say $3$ or $5$. Equation~(\ref{eq:one}) then implies that neither $f_{33}$ nor $\beta_2$ can be too small either.  Hereafter we will take $f_{33} \sim 1$ and $\beta_2 \sim 1$ (or as we shall see, one of the $\beta_i \sim 1$), and note that they can have  slightly different values that also generate $m_e$, and this doesn't change the crux of our results. 

The above parameters determine the P breaking or $SU(2)_L \times U(1)_{B-L}$ breaking scale from equation~(\ref{eq:vL}) and the first term of equation~(\ref{eq:mnu}),  and the observed light neutrino mass scale $\sim 0.05 eV$. Therefore the generation of electron mass through the $\beta_2$ coupling implies
\begin{equation}
    \sqrt{\Delta m^2_{23}} \sim 0.05 eV \sim f_{33}\beta_2 v_{wk}^2/(\rho_3-2\rho_1)v_R
    \end{equation}

 Substituting $f_{33}, \beta_2 \sim 1$  we get $v_R \sim 10^{14-15} GeV $.  We have set $\rho_3, \rho_1 \sim 1$  in the denominator as they anyway get generated in one loop RGE.

% It is interesting that we have obtained $v_R$ without using any grand unification constraints.
 
The beta function of $\tilde{h}^\ell$,  $\beta^{(2)}_{\tilde{h}^\ell} \sim (6\beta_1/16\pi^2)f^\dagger h^{\ell^\star}f$ contributes to $h^-$ in eq.~(\ref{eq:diracyuk}) (through $\tilde{h}^\ell \kappa_1$ term whose diagonal entries at scale $M'$ we take to be the mu and tau Yukawa couplings),  and the $m_e$ generated on RGE running in analogy with equation~(\ref{eq:me}) is 
\begin{equation}
%h^D v_{wk}{array}{ll}
 m_e \approx {{6  (h^\ell_{33} \kappa_1)} \over {(16 \pi^2)^2}} c'_{\nu mix} \left( f_{33}^2 \beta_1\right) ln(M'/v_R)  
 \label{eq:beta1mass}
 %\end{array}
\end{equation}
with $c'_{\nu mix} \sim c_{\nu mix}$. If the above generates $m_e$, then $v_R$ can be lowered -- if we take $f_{33}, \beta_1 \approx 1$ then $h^\ell_{33}$ must be $\sim 0.01$ (near the tau Yukawa) to get the correct value for $m_e$.   Since the second term in equation~(\ref{eq:mnu}) also contributes to the light neutrino mass, the lower bound on $v_R$ is a factor $h^{\ell^{2}}_{33}$ lower than what we obtained earlier, or $v_R \gtrsim 10^{11}GeV$.  If $v_R$ is lowered more by further lowering $h^\ell_{33}$, then sufficient mass for the electron is not generated via $\beta_1$ coupling.

% Therefore in our model we predict 
% \begin{equation}
%     v_R \sim 10^{11}~to~10^{15} GeV.
%     \label{eq:vR}
 %\end{equation}

Our above results %(including those in Table~\ref{tab:vR}) 
that couplings $f_{33}$ (that sets the scale for $f_{ij}$) and one of the $\beta_i$ are $\sim \mathcal{O}(1) $ (or sufficiently large), 
follow from sufficient electron mass being generated in our model, and we have not a priori assumed that these must be large or small. Having  obtained these, just to appreciate the possible physics, we consider the approximate $\mu_Q-$symmetry  %To appreciate the physics consider the approximately broken $\mu_Q-$symmetry under which
under which
 \begin{equation}
     \phi \rightarrow e^{i\alpha}\phi \ ( \mathrel{\mkern-3mu\Rightarrow\mkern-3mu}  \tilde{\phi} \rightarrow e^{-i\alpha} \tilde{\phi}), \  Q_{iL} \rightarrow e^{i\alpha}Q_{iL}
     \label{eq:muq}
 \end{equation}
 so that the resulting top quark's SM Yukawa is unsuppressed, while the bottom quark's Yukawa coupling $h_b$ (as also the VEV $\kappa_2$)  occurs from terms that break $\mu_Q$ (such as $\mu_2^2 Tr(\tilde{\phi}^\dagger \phi)$) and is therefore suppressed, $h_b \sim 1/40$.  

Since $h^\ell$ as well as $\tilde{h}^\ell$ terms in equation~\ref{eq:hyuk} break the above $\mu_Q-$symmetry,  we would expect both $h^\ell_{33}$ and $\tilde{h}^\ell_{33}$ (and therefore also $h^D_{33}$) be suppressed $\sim h_b$ or $h_\tau$. Moreover from equation~(\ref{eq:beta2}),  $\beta_1$ respects the $\mu_Q-$symmetry, while $\beta_2$ and $\beta_3$ don't and are suppressed.   This provides the physical context for electron mass generation from equation~(\ref{eq:beta1mass}) via the unsuppressed $\beta_1 \approx 1$ coupling, slightly suppressed $h^\ell_{33}, \tilde{h}^\ell_{33}$,  and  the range of $v_R$ scale shown in the  $\beta_1 \approx 1$ column of  Table~(\ref{tab:vR}).

\begin{table}[t]
\centering
\begin{tabular}{lccc}
\hline 
& 
$\beta_1 \approx 1$
&
$\beta_2 \approx 1$
&
$\beta_3 \approx 1$
 \\ 
 \hline 
 $\ \ v_R \ \ $
 &
 $\ \ 10^{12-14} GeV \ \ $
 &
  $\ \ 10^{14-15} GeV \ \ $
&
$ \ \ 10^{10-13} GeV \ \ $
\\ \hline

\end{tabular}
\caption{Range for $U(1)_{B-L}$ breaking scale  $v_R$ obtained from~(\ref{eq:vL}) and first term in~(\ref{eq:mnu})  using $\kappa_1 \approx v_{wk}, \kappa_2/\kappa_1 \sim 0.1~to~10^{-2.5},  \rho_3 \sim 1, \sqrt{\Delta m_{23}^2} \sim 0.05 eV $. To generate $m_e$: $f_{33} \sim 1$ and one of the $\beta_i\approx 1$. To obtain $v_R \sim 10 TeV$:  $\beta_3 \approx 1$, and $\kappa_2/\kappa_1 \sim 10^{-5.5}$ which needs fine-tuning. }
\label{tab:vR}
\end{table}

On the other hand, instead of $\mu_Q$, if  we have an approximate $\mu_{QL}$ symmetry under which,
\begin{equation}
\phi, Q_{iL}, L_{iL}\rightarrow e^{i\alpha} \{ \phi, Q_{iL}, L_{iL} \}; \ \Delta_L, \Delta'_L \rightarrow e^{-2i\alpha}\{\Delta_L,\Delta'_L\}
\label{eq:muql}
\end{equation}
then $\beta_2, f$ and $h^\ell$ respect the symmetry, while $\beta_1, \beta_3, \tilde{h}^\ell$ break the symmetry and are suppressed.  This provides the context for $m_e$ generation through $\beta_2 \sim 1, h^\ell_{33} \kappa_2$ in equations~(\ref{eq:betatwoloop}),(\ref{eq:me}) and~(\ref{eq:masselec}). In this case there can also be a similar contribution  from a suppressed $\beta_1 \sim 0.01$ and unsuppressed $h^\ell_{33}\kappa_1$ through equation~(\ref{eq:beta1mass}). 

The electron mass can also be generated if $\beta_3 f^2_{33} \approx 1$, using the beta function for $\tilde{h}^\ell$ obtained by replacing $\beta_2 \rightarrow \beta_3$ and $h^\ell \rightarrow \tilde{h}^\ell$ on both sides of equation~(\ref{eq:betatwoloop}). This leads to equation~(\ref{eq:me}) with $\beta_2 \rightarrow \beta_3$.  In this case, if we set $\beta_1 \approx \beta_2 \approx 0$, then we can further lower the scale of $v_R$  calculated using equations~(\ref{eq:vL}) and (\ref{eq:mnu}) since $|\kappa_2|/\kappa_1,  h^\ell_{33}$ (and therefore also $h^D_{33}$) can be chosen to be $\sim 10^{-5.5}$ (so that $v_R \sim 10 TeV$).  However since $|\kappa_2| \gtrsim 10^{-2.5}\kappa_1$ due to tree-level or one loop RGE generation, % of Higgs potential terms that contains $Tr \tilde{\phi}^\dagger \phi$, 
it requires fine-tuned cancellations to lower $\kappa_2$ for obtaining $v_R \sim 10TeV$, whose  natural scale is given in Table~\ref{tab:vR}. 

In order to obtain $v_R \sim 1-10 TeV$ without ``heavy fine-tuning" (quoting~\cite{Dev2016}) P can be broken softly so that $\Delta_L$ gets a much heavier mass $>> v_R$\cite{Dev2016, PhysRevLett.52.1072}.  Even now, we can generate $m_e$ by RGE running above the $\Delta_L$ mass.   In this case Table~\ref{tab:vR} will not apply.
%using $\tilde{h}_{33} \sim 0.01$ and $\beta_i \sim 1$, and based on $\Delta_L$ mass use either Type I or Type II seesaw.  

\section{Muon mass}
\label{sec:mu}
%\textit{Muon mass.} 
If we use $\chi-$symmetry (like eq.~(\ref{eq:chidis})) to set both the electron and muon Yukawa couplings to zero, and break it softly at scale $M'$ by the mixing of $\Delta'_{L,R}$ and $\Delta_{L,R}$ fields as before, and take both $\beta_1f^2_{33}, h^\ell_{33} \sim 1 $, then the mass of the muon (rather than the electron) is generated by  the right hand side of equation~(\ref{eq:beta1mass}).  Since originally the electron and the muon masses are both zero and degenerate,  sufficient mass for the electron is not generated efficiently at the two loop level. However if we add triplets $\Delta''_{L,R}$ at scale $\sim M'$, then the threshold effects due to differing masses at this scale can first lift the degeneracy and then generate both the muon and electron masses with $\beta_1, f_{33}, h^\ell_{33} \sim 1$ (and therefore $v_R \sim 10^{14-15} GeV)$. The tau mass is $\sim h^\ell_{33}\kappa_2$, so that all lepton masses are generated via $\mathcal{O}(1)$ Yukawa parameters. 

\section{Testing the model}
\label{sec:test}
%\textit{Testing the model.} 
P sets $\theta_{QCD}$ to zero and the strong CP phase $\bar{\theta}$ generated 
%at scale $v_R$
on its spontaneous breaking depends on the imaginary part of a dimensionless quartic coupling ($\alpha_2$) of the Higgs potential of the minimal left right symmetric model. %that generates an imaginary component to the VEV $\kappa_2$ that contributes to the bottom quark's mass via the top quark's Yukawa coupling.  
As  was shown in reference~\cite{Kuchimanchi_2015}, leptons circulating in one internal loop contribute through the Yukawa matrices $h^\ell, \tilde{h}^\ell$ and $f$ to $\alpha_2$ so that  
% the CP violation in leptonic Yukawa couplings, on one loop RGE running,  generates this imaginary part so that
\begin{equation}
\bar{\theta} \sim \left(1/16 \pi^2\right)(m_t/m_b)\left|Tr\left(f^\dagger f [h^\ell, \tilde{h}^\ell]\right)\right| \ ln(M'/v_R)
\label{eq:theta-rge}
\end{equation}
is generated along with the electron mass on RGE running between $M'$ and $v_R$ scales. The trace in the above is purely imaginary 
%(and would be zero if there is no leptonic CP violation) 
since the commutator $[h^\ell, \tilde{h}^\ell]$ is anti-Hermitian, and top and bottom quark mass ratio $m_{t}/m_b \sim 40$. %The above is the magnitude of $\bar{\theta}$ generated and its sign can be plus or minus. 
Importantly, the dependence on $v_R$ is logarithmic and  the contribution is significant even for $v_R \sim  10^{10-15} GeV$, which provides the way to test our model.

Since to generate the electron's mass   $f_{ij} \sim f_{33} \sim 1$ (due to equations such as~(\ref{eq:one})), and to generate the tau mass one of the Dirac type Yukawa couplings, say $\tilde{h}^\ell_{33} \gtrsim 10^{-2}$, we obtain from eq.~(\ref{eq:theta-rge}),
\begin{equation}
    |\bar{\theta}| \sim f_{33}^2 \tilde{h}^\ell_{33} h^\ell_{33} |sin\delta_{CP}|/4 \sim 10^{-2.5} h^\ell_{33}|sin\delta_{CP}|
    \label{eq:theta}
\end{equation}
where we have taken in eq~(\ref{eq:theta-rge}), $ln{(M'/v_R)} \sim 10$, and $[h^\ell, \tilde{h}^\ell] \sim (\tilde{h}^\ell_{33}) (h^\ell_{33})/10$, as its off diagonal terms can have a tree-level value and also get generated in RGE running from $f_{ij} \sim 1$. $\delta_{CP}$ is the Dirac CP violating phase in the PMNS matrix obtained from CP phases present in $f, h^\ell, \tilde{h}^\ell$. 

$|\bar{\theta}| \leq 10^{-10}$  from nEDM experiments~\cite{PhysRevLett.124.081803}  implies due to eq.~(\ref{eq:theta}) that for $h^\ell_{33} \geq 10^{-5.5}$, $\delta_{CP} \leq 0.01$ mod $\pi$.  Therefore for $10 TeV \leq v_R \leq 10^{15}GeV$ (see Table~\ref{tab:vR}), we expect $|sin \delta_{CP}| \leq 0.01$.

For $h^\ell_{33} \gtrsim 10^{-2}$, which is the  likely value of suppression due to approximate $\mu_Q$ or $\mu_{QL}$ symmetries (eqs.~({\ref{eq:muq}}) and ~(\ref{eq:muql})), we obtain even smaller $\delta_{CP} \leq 10^{-5.5}$ (mod $\pi$).

Thus, with the exception of a highly tuned region where $h^\ell_{33} < 10^{-6}$, we expect that upcoming neutrino experiments Hyper-K~\cite{10.3389/fphy.2024.1378254} and DUNE~\cite{universe10050221} will make a discovery consistent with the absence of CP violation in the leptonic PMNS matrix. 

As of now, joint analysis~\cite{zoya} of NOvA and T2K experiments reveals $\delta_{CP} = 203^{+63}_{-37}$, while the global best fit of neutrino experiments by Nu-Fit 5.3 (2024)~\cite{Esteban_2020} is $\delta_{CP}=197^{+41}_{-25}$,  for normal ordering of neutrino masses. Our expectation of absence of leptonic CP violation,  $\delta_{CP} = 0^o$ or $180^o$ ($0$ mod $\pi$), is currently within one sigma. %  It is exciting to have a model for radiative generation of electron (and also muon) mass that is testable, despite its high energy scale. 

Interestingly after the latest data release and analysis, Nu-Fit 6.0 (2024)~\cite{esteban2024nufit60updatedglobalanalysis} has updated the above value to $\delta_{CP} = 177^{+19}_{-20}$ (for normal ordering with $\theta_{23}$ in upper octant. And the normal ordering, lower octant fit is also consistent with $\delta_{CP} = 180^o$ to within one sigma).
   
Before moving to the next section we also note that as stated in~\cite{Kuchimanchi_2015}, $|\bar{\theta}|$ generated from tree-level and one-loop contribution to $Im(\alpha_2)$ are independently $\leq 10^{-10}$ as we do not allow fine tuned cancellations between them.
The strong CP problem can also be solved~\cite{Kuchimanchi:2010xs, PhysRevD.108.095023} so that there is no tree-level contribution.

\section{Charge Conjugation, Unification.}
\label{sec:C}
%That leptonic CP violation must be absent is a test for the discrete space time symmetry of P~\cite{PhysRevD.108.095023}, but if there is grand unification we would expect it to be present.
P beautifully controls the strong CP problem and the absence of leptonic CP violating phases is a way to test it (the minimal extension that solves the strong CP problem~\cite{Kuchimanchi:2010xs} also predicts an absence of leptonic CP violation~\cite{Kuchimanchi:2012te,PhysRevD.108.095023}). 

But if there is grand unification we'd expect CP violation. However this does not cause a conflict because it is the left right symmetric model with C (rather than P) that unifies to $SO(10)$.  If we use C  to exchange the $SU(2)_L$ and $SU(2)_R$ groups, the same results for the electron mass and $v_R$ scale follow.  However with C (and no P), many other CP phases are present and they as well as the CKM phase can generate $\bar{\theta}$ at tree level and via RGE running.  Hence we would likely introduce axions~\cite{1977PhRvL..38.1440P, PhysRevLett.40.223,PhysRevLett.40.279} anyway, in which case there is no prediction of the absence of leptonic $\delta_{CP}$.  % Thus the direction of C and $SO(10)$ unification can be tested in the usual manner via proton decay, discovery of supersymmetry for SUSY GUTS, and  axions to control the strong CP problem, while the direction of discrete space time symmetries can be tested via the absence of leptonic CP violation.        

\section{Gauging the $\chi$ symmetry}
\label{sec:gauge}
We can break the discrete $\chi$ symmetry in eqn.~(\ref{eq:chidis}) spontaneously by adding  a real scalar singlet $\sigma_L$ ($\rightarrow -\sigma_L$ under $\chi$ symmetry of eqn.~(\ref{eq:chidis})) and its parity partner $\sigma_R$ which is even under $\chi$. $\left<\sigma_L\right> = \left<\sigma_R\right> \sim M'> v_R$ breaks $\chi-$symmetry above the $P$ breaking scale and generates $\mu'^2$ in eqn.~(\ref{eq:vmass}) spontaneously from the $\chi$ invariant Higgs potential term $\mu\sigma_L Tr(\Delta'^\dagger_L \Delta_L) + h.c. + L \rightarrow R.$

If $\chi$ is a continuous $U(1)_L$ symmetry as in eqn.~(\ref{eq:chi}), then it can be gauged  by adding a complex $SU(2)_L \times SU(2)_R$ singlet scalar $\sigma_L$ (with $B-L = 0$) and a heavy vector like lepton whose left and right components $L_{4L}, L'_R$ are $SU(2)_L$ doublets with $B-L = -1$.  Under the transformation of eqn.~(\ref{eq:chi}), $L'_R, \sigma_L \rightarrow e^{i\chi} \{L'_R, \sigma_L\}$ so that all the anomalies cancel and the $U(1)_L$ symmetry is gauged. 

Because of $P$, there is a  corresponding $U(1)_R$ symmetry with singlet $\sigma_R$  (under $P$, $\sigma_L \leftrightarrow \sigma_R$, $U(1)_L \leftrightarrow U(1)_R$) 
and $SU(2)_R$ doublets $L_{4R}, L'_L$, so that the gauge symmetry $U(1)_L \times U(1)_R$ is broken spontaneously above the $P$ breaking scale $v_R$ by the VEVs $\left<\sigma_L\right> = \left<\sigma_R\right> \sim M' > v_R$. Note that $U(1)_R$  charges have been assigned to $L_{1R}, \Delta'_R, L'_L$ and $\sigma_R$ analogous to $U(1)_L$ charges of their parity counterparts. The $\mu'^2$ in eqn.~(\ref{eq:vmass}) is now spontaneously generated from the gauge invariant Higgs potential coupling $\mu\sigma^\star_L Tr(\Delta'^\dagger_L \Delta_L) + h.c.+ L \rightarrow R$. And the heavy vector-like leptons pick up a mass due to  Yukawa coupling $h^\ell_\sigma\sigma^\star_L \bar{L}_{4L} L'_R + L \leftrightarrow R$. %, where  $h_\sigma$ can be small or large. %, so that the heavy leptons can have masses from the weak/TeV scale to $M'$. 
%Note that $P$ interchanges the gauge bosons of $U(1)_L$ and $U(1)_R$.

We assign an intrinsic party $\eta = i$ to the heavy leptons so that under $P:$  $L_{4L,4R}; L'_{L,R} \rightarrow i \{L_{4R,4L}; L'_{R,L}\}$, and thereby mixing terms between heavy and usual leptons (whose $\eta = 1$) such as  $\bar{L}_{1L} L'_R, \bar{L}_{1R} L'_L, \sigma^\star_L \bar{L}_{3L} L'_R $ are absent due to $P$~\cite{Kuchimanchi:2012te}. The electron is massless at the tree level at scale $M'$, and its mass is generated in the two loop RGE running discussed in this work.  

The Dirac and Majorana Yukawa terms $h^\ell_{44}\bar{L}_{4L} \phi L_{4R}$, $\tilde{h}^\ell_{44}\bar{L}_{4L} \tilde{\phi} L_{4R}$  and $if_{44} (L^T_{4L} \tau_2 \Delta_L L_{4L} - L^T_{4R} \tau_2 \Delta_R L_{4R})$ (note the minus sign due to the above intrinsic parity $\eta = i$) are present, with $h^\ell_{44}$ and $\tilde{h}^\ell_{44}$ real due to P.   

Such heavy vector-like leptons with $\eta = i$ have been considered before in the context of strong CP problem and dark matter~\cite{Kuchimanchi:2012te,PhysRevD.108.095023}.  Interestingly, there are efforts to also gauge P~\cite{Maiezza:2021dui}.

\section{Conclusion.} 
\label{sec:conc}
We completed the minimal left right symmetric model in the ultraviolet by restoring a discrete or a continuous $U(1)_L \times U(1)_R$ chiral symmetry (which can be gauged)  so that we have a  predictive theory where the electron's mass is radiatively generated from its neutrino mixing in two loop RGE running of the minimal model, $B-L$ gauge symmetry breaking scale is predicted to be $10^{10} GeV \lesssim v_R \leq 10^{15} GeV$ with Type 2 seesaw mechanism,  and because of the  RGE contribution from leptonic CP phases to the strong CP phase,  the model is testable by neutrino experiments DUNE and Hyper-K for which we anticipate a null result - the non-discovery of leptonic CP violation in the Dirac CP phase of the PMNS matrix ($\delta_{CP} = 0$ or $180^o$ to within a degree).   If there are axions then there is no prediction of absence of leptonic $\delta_{CP}$. We also anticipate normal ordering of neutrino masses with $U(1)_L \times U(1)_R$ and the scalar content of this work.

\section*{Acknowledgment} Thanks to Kaustubh Agashe for insightful discussions and helpful suggestions.
 
%\raggedright % Left-justify the bibliography entries
%\bibliographystyle{apsrev4-2} % Specify the bibliography style			
\bibliography{mass-heir}

%apsrev4-2.bst 2019-01-14 (MD) hand-edited version of apsrev4-1.bst
%Control: key (0)
%Control: author (8) initials jnrlst
%Control: editor formatted (1) identically to author
%Control: production of article title (0) allowed
%Control: page (0) single
%Control: year (1) truncated
%Control: production of eprint (0) enabled
\begin{thebibliography}{39}%
\makeatletter
\providecommand \@ifxundefined [1]{%
 \@ifx{#1\undefined}
}%
\providecommand \@ifnum [1]{%
 \ifnum #1\expandafter \@firstoftwo
 \else \expandafter \@secondoftwo
 \fi
}%
\providecommand \@ifx [1]{%
 \ifx #1\expandafter \@firstoftwo
 \else \expandafter \@secondoftwo
 \fi
}%
\providecommand \natexlab [1]{#1}%
\providecommand \enquote  [1]{``#1''}%
\providecommand \bibnamefont  [1]{#1}%
\providecommand \bibfnamefont [1]{#1}%
\providecommand \citenamefont [1]{#1}%
\providecommand \href@noop [0]{\@secondoftwo}%
\providecommand \href [0]{\begingroup \@sanitize@url \@href}%
\providecommand \@href[1]{\@@startlink{#1}\@@href}%
\providecommand \@@href[1]{\endgroup#1\@@endlink}%
\providecommand \@sanitize@url [0]{\catcode `\\12\catcode `\$12\catcode `\&12\catcode `\#12\catcode `\^12\catcode `\_12\catcode `\%12\relax}%
\providecommand \@@startlink[1]{}%
\providecommand \@@endlink[0]{}%
\providecommand \url  [0]{\begingroup\@sanitize@url \@url }%
\providecommand \@url [1]{\endgroup\@href {#1}{\urlprefix }}%
\providecommand \urlprefix  [0]{URL }%
\providecommand \Eprint [0]{\href }%
\providecommand \doibase [0]{https://doi.org/}%
\providecommand \selectlanguage [0]{\@gobble}%
\providecommand \bibinfo  [0]{\@secondoftwo}%
\providecommand \bibfield  [0]{\@secondoftwo}%
\providecommand \translation [1]{[#1]}%
\providecommand \BibitemOpen [0]{}%
\providecommand \bibitemStop [0]{}%
\providecommand \bibitemNoStop [0]{.\EOS\space}%
\providecommand \EOS [0]{\spacefactor3000\relax}%
\providecommand \BibitemShut  [1]{\csname bibitem#1\endcsname}%
\let\auto@bib@innerbib\@empty
%</preamble>
\bibitem [{\citenamefont {Fritzsch}(1978)}]{FRITZSCH1978317}%
  \BibitemOpen
  \bibfield  {author} {\bibinfo {author} {\bibfnamefont {H.}~\bibnamefont {Fritzsch}},\ }\bibfield  {title} {\bibinfo {title} {Weak-interaction mixing in the six-quark theory},\ }\href {https://doi.org/https://doi.org/10.1016/0370-2693(78)90524-5} {\bibfield  {journal} {\bibinfo  {journal} {Physics Letters B}\ }\textbf {\bibinfo {volume} {73}},\ \bibinfo {pages} {317} (\bibinfo {year} {1978})}\BibitemShut {NoStop}%
\bibitem [{\citenamefont {Altarelli}\ \emph {et~al.}(2000)\citenamefont {Altarelli}, \citenamefont {Feruglio},\ and\ \citenamefont {Masina}}]{Altarelli_2000}%
  \BibitemOpen
  \bibfield  {author} {\bibinfo {author} {\bibfnamefont {G.}~\bibnamefont {Altarelli}}, \bibinfo {author} {\bibfnamefont {F.}~\bibnamefont {Feruglio}},\ and\ \bibinfo {author} {\bibfnamefont {I.}~\bibnamefont {Masina}},\ }\bibfield  {title} {\bibinfo {title} {Large neutrino mixing from small quark and lepton mixings},\ }\href {https://doi.org/10.1016/s0370-2693(99)01360-x} {\bibfield  {journal} {\bibinfo  {journal} {Physics Letters B}\ }\textbf {\bibinfo {volume} {472}},\ \bibinfo {pages} {382–391} (\bibinfo {year} {2000})}\BibitemShut {NoStop}%
\bibitem [{\citenamefont {Hall}\ \emph {et~al.}(2000)\citenamefont {Hall}, \citenamefont {Murayama},\ and\ \citenamefont {Weiner}}]{PhysRevLett.84.2572}%
  \BibitemOpen
  \bibfield  {author} {\bibinfo {author} {\bibfnamefont {L.}~\bibnamefont {Hall}}, \bibinfo {author} {\bibfnamefont {H.}~\bibnamefont {Murayama}},\ and\ \bibinfo {author} {\bibfnamefont {N.}~\bibnamefont {Weiner}},\ }\bibfield  {title} {\bibinfo {title} {Neutrino mass anarchy},\ }\href {https://doi.org/10.1103/PhysRevLett.84.2572} {\bibfield  {journal} {\bibinfo  {journal} {Phys. Rev. Lett.}\ }\textbf {\bibinfo {volume} {84}},\ \bibinfo {pages} {2572} (\bibinfo {year} {2000})}\BibitemShut {NoStop}%
\bibitem [{\citenamefont {Haba}\ and\ \citenamefont {Murayama}(2001)}]{PhysRevD.63.053010}%
  \BibitemOpen
  \bibfield  {author} {\bibinfo {author} {\bibfnamefont {N.}~\bibnamefont {Haba}}\ and\ \bibinfo {author} {\bibfnamefont {H.}~\bibnamefont {Murayama}},\ }\bibfield  {title} {\bibinfo {title} {Anarchy and hierarchy: An approach to study models of fermion masses and mixings},\ }\href {https://doi.org/10.1103/PhysRevD.63.053010} {\bibfield  {journal} {\bibinfo  {journal} {Phys. Rev. D}\ }\textbf {\bibinfo {volume} {63}},\ \bibinfo {pages} {053010} (\bibinfo {year} {2001})}\BibitemShut {NoStop}%
\bibitem [{\citenamefont {Weinberg}(1980)}]{PhysRevD.22.1694}%
  \BibitemOpen
  \bibfield  {author} {\bibinfo {author} {\bibfnamefont {S.}~\bibnamefont {Weinberg}},\ }\bibfield  {title} {\bibinfo {title} {Varieties of baryon and lepton nonconservation},\ }\href {https://doi.org/10.1103/PhysRevD.22.1694} {\bibfield  {journal} {\bibinfo  {journal} {Phys. Rev. D}\ }\textbf {\bibinfo {volume} {22}},\ \bibinfo {pages} {1694} (\bibinfo {year} {1980})}\BibitemShut {NoStop}%
\bibitem [{\citenamefont {Altarelli}\ and\ \citenamefont {Feruglio}(2004)}]{Altarelli_2004}%
  \BibitemOpen
  \bibfield  {author} {\bibinfo {author} {\bibfnamefont {G.}~\bibnamefont {Altarelli}}\ and\ \bibinfo {author} {\bibfnamefont {F.}~\bibnamefont {Feruglio}},\ }\bibfield  {title} {\bibinfo {title} {Models of neutrino masses and mixings},\ }\href {https://doi.org/10.1088/1367-2630/6/1/106} {\bibfield  {journal} {\bibinfo  {journal} {New Journal of Physics}\ }\textbf {\bibinfo {volume} {6}},\ \bibinfo {pages} {106} (\bibinfo {year} {2004})}\BibitemShut {NoStop}%
\bibitem [{\citenamefont {Bonilla}\ \emph {et~al.}(2015)\citenamefont {Bonilla}, \citenamefont {Fonseca},\ and\ \citenamefont {Valle}}]{Bonilla:2015eha}%
  \BibitemOpen
  \bibfield  {author} {\bibinfo {author} {\bibfnamefont {C.}~\bibnamefont {Bonilla}}, \bibinfo {author} {\bibfnamefont {R.~M.}\ \bibnamefont {Fonseca}},\ and\ \bibinfo {author} {\bibfnamefont {J.~W.~F.}\ \bibnamefont {Valle}},\ }\bibfield  {title} {\bibinfo {title} {{Consistency of the triplet seesaw model revisited}},\ }\href {https://doi.org/10.1103/PhysRevD.92.075028} {\bibfield  {journal} {\bibinfo  {journal} {Phys. Rev. D}\ }\textbf {\bibinfo {volume} {92}},\ \bibinfo {pages} {075028} (\bibinfo {year} {2015})},\ \Eprint {https://arxiv.org/abs/1508.02323} {arXiv:1508.02323 [hep-ph]} \BibitemShut {NoStop}%
\bibitem [{\citenamefont {Pati}\ and\ \citenamefont {Salam}(1974)}]{PhysRevD.10.275}%
  \BibitemOpen
  \bibfield  {author} {\bibinfo {author} {\bibfnamefont {J.~C.}\ \bibnamefont {Pati}}\ and\ \bibinfo {author} {\bibfnamefont {A.}~\bibnamefont {Salam}},\ }\bibfield  {title} {\bibinfo {title} {Lepton number as the fourth "color"},\ }\href {https://doi.org/10.1103/PhysRevD.10.275} {\bibfield  {journal} {\bibinfo  {journal} {Phys. Rev. D}\ }\textbf {\bibinfo {volume} {10}},\ \bibinfo {pages} {275} (\bibinfo {year} {1974})}\BibitemShut {NoStop}%
\bibitem [{\citenamefont {Mohapatra}\ and\ \citenamefont {Pati}(1975)}]{PhysRevD.11.566}%
  \BibitemOpen
  \bibfield  {author} {\bibinfo {author} {\bibfnamefont {R.~N.}\ \bibnamefont {Mohapatra}}\ and\ \bibinfo {author} {\bibfnamefont {J.~C.}\ \bibnamefont {Pati}},\ }\bibfield  {title} {\bibinfo {title} {Left-right gauge symmetry and an "isoconjugate" model of $cp$ violation},\ }\href {https://doi.org/10.1103/PhysRevD.11.566} {\bibfield  {journal} {\bibinfo  {journal} {Phys. Rev. D}\ }\textbf {\bibinfo {volume} {11}},\ \bibinfo {pages} {566} (\bibinfo {year} {1975})}\BibitemShut {NoStop}%
\bibitem [{\citenamefont {Senjanovic}\ and\ \citenamefont {Mohapatra}(1975)}]{Senjanovic:1975rk}%
  \BibitemOpen
  \bibfield  {author} {\bibinfo {author} {\bibfnamefont {G.}~\bibnamefont {Senjanovic}}\ and\ \bibinfo {author} {\bibfnamefont {R.~N.}\ \bibnamefont {Mohapatra}},\ }\bibfield  {title} {\bibinfo {title} {{Exact Left-Right Symmetry and Spontaneous Violation of Parity}},\ }\href {https://doi.org/10.1103/PhysRevD.12.1502} {\bibfield  {journal} {\bibinfo  {journal} {Phys. Rev.}\ }\textbf {\bibinfo {volume} {D12}},\ \bibinfo {pages} {1502} (\bibinfo {year} {1975})}\BibitemShut {NoStop}%
%%CITATION = PHRVA,D12,1502;%%
\bibitem [{\citenamefont {Deshpande}\ \emph {et~al.}(1991)\citenamefont {Deshpande}, \citenamefont {Gunion}, \citenamefont {Kayser},\ and\ \citenamefont {Olness}}]{PhysRevD.44.837}%
  \BibitemOpen
  \bibfield  {author} {\bibinfo {author} {\bibfnamefont {N.~G.}\ \bibnamefont {Deshpande}}, \bibinfo {author} {\bibfnamefont {J.~F.}\ \bibnamefont {Gunion}}, \bibinfo {author} {\bibfnamefont {B.}~\bibnamefont {Kayser}},\ and\ \bibinfo {author} {\bibfnamefont {F.}~\bibnamefont {Olness}},\ }\bibfield  {title} {\bibinfo {title} {Left-right-symmetric electroweak models with triplet higgs field},\ }\href {https://doi.org/10.1103/PhysRevD.44.837} {\bibfield  {journal} {\bibinfo  {journal} {Phys. Rev. D}\ }\textbf {\bibinfo {volume} {44}},\ \bibinfo {pages} {837} (\bibinfo {year} {1991})}\BibitemShut {NoStop}%
\bibitem [{\citenamefont {Duka}\ \emph {et~al.}(2000)\citenamefont {Duka}, \citenamefont {Gluza},\ and\ \citenamefont {Zralek}}]{Duka:1999uc}%
  \BibitemOpen
  \bibfield  {author} {\bibinfo {author} {\bibfnamefont {P.}~\bibnamefont {Duka}}, \bibinfo {author} {\bibfnamefont {J.}~\bibnamefont {Gluza}},\ and\ \bibinfo {author} {\bibfnamefont {M.}~\bibnamefont {Zralek}},\ }\bibfield  {title} {\bibinfo {title} {{Quantization and renormalization of the manifest left- right symmetric model of electroweak interactions}},\ }\href {https://doi.org/10.1006/aphy.1999.5988} {\bibfield  {journal} {\bibinfo  {journal} {Annals Phys.}\ }\textbf {\bibinfo {volume} {280}},\ \bibinfo {pages} {336} (\bibinfo {year} {2000})},\ \Eprint {https://arxiv.org/abs/hep-ph/9910279} {arXiv:hep-ph/9910279} \BibitemShut {NoStop}%
%%CITATION = HEP-PH/9910279;%%
\bibitem [{\citenamefont {Hooft}(1980)}]{Hooft1980}%
  \BibitemOpen
  \bibfield  {author} {\bibinfo {author} {\bibfnamefont {G.}~\bibnamefont {Hooft}},\ }\bibinfo {title} {Naturalness, chiral symmetry, and spontaneous chiral symmetry breaking},\ in\ \href {https://doi.org/10.1007/978-1-4684-7571-5_9} {\emph {\bibinfo {booktitle} {Recent Developments in Gauge Theories}}},\ \bibinfo {editor} {edited by\ \bibinfo {editor} {\bibfnamefont {G.}~\bibnamefont {Hooft}}, \bibinfo {editor} {\bibfnamefont {C.}~\bibnamefont {Itzykson}}, \bibinfo {editor} {\bibfnamefont {A.}~\bibnamefont {Jaffe}}, \bibinfo {editor} {\bibfnamefont {H.}~\bibnamefont {Lehmann}}, \bibinfo {editor} {\bibfnamefont {P.~K.}\ \bibnamefont {Mitter}}, \bibinfo {editor} {\bibfnamefont {I.~M.}\ \bibnamefont {Singer}},\ and\ \bibinfo {editor} {\bibfnamefont {R.}~\bibnamefont {Stora}}}\ (\bibinfo  {publisher} {Springer US},\ \bibinfo {address} {Boston, MA},\ \bibinfo {year} {1980})\ pp.\ \bibinfo {pages} {135--157}\BibitemShut {NoStop}%
\bibitem [{\citenamefont {Balakrishna}(1988{\natexlab{a}})}]{Balakrishna:1987qd}%
  \BibitemOpen
  \bibfield  {author} {\bibinfo {author} {\bibfnamefont {B.~S.}\ \bibnamefont {Balakrishna}},\ }\bibfield  {title} {\bibinfo {title} {{Fermion Mass Hierarchy From Radiative Corrections}},\ }\href {https://doi.org/10.1103/PhysRevLett.60.1602} {\bibfield  {journal} {\bibinfo  {journal} {Phys. Rev. Lett.}\ }\textbf {\bibinfo {volume} {60}},\ \bibinfo {pages} {1602} (\bibinfo {year} {1988}{\natexlab{a}})}\BibitemShut {NoStop}%
\bibitem [{\citenamefont {Balakrishna}\ \emph {et~al.}(1988)\citenamefont {Balakrishna}, \citenamefont {Kagan},\ and\ \citenamefont {Mohapatra}}]{Balakrishna:1988ks}%
  \BibitemOpen
  \bibfield  {author} {\bibinfo {author} {\bibfnamefont {B.~S.}\ \bibnamefont {Balakrishna}}, \bibinfo {author} {\bibfnamefont {A.~L.}\ \bibnamefont {Kagan}},\ and\ \bibinfo {author} {\bibfnamefont {R.~N.}\ \bibnamefont {Mohapatra}},\ }\bibfield  {title} {\bibinfo {title} {{Quark Mixings and Mass Hierarchy From Radiative Corrections}},\ }\href {https://doi.org/10.1016/0370-2693(88)91676-0} {\bibfield  {journal} {\bibinfo  {journal} {Phys. Lett. B}\ }\textbf {\bibinfo {volume} {205}},\ \bibinfo {pages} {345} (\bibinfo {year} {1988})}\BibitemShut {NoStop}%
\bibitem [{\citenamefont {Balakrishna}(1988{\natexlab{b}})}]{Balakrishna:1988xg}%
  \BibitemOpen
  \bibfield  {author} {\bibinfo {author} {\bibfnamefont {B.~S.}\ \bibnamefont {Balakrishna}},\ }\bibfield  {title} {\bibinfo {title} {{RADIATIVELY INDUCED LEPTON MASSES}},\ }\href {https://doi.org/10.1016/0370-2693(88)91480-3} {\bibfield  {journal} {\bibinfo  {journal} {Phys. Lett. B}\ }\textbf {\bibinfo {volume} {214}},\ \bibinfo {pages} {267} (\bibinfo {year} {1988}{\natexlab{b}})}\BibitemShut {NoStop}%
\bibitem [{\citenamefont {Barr}(1990)}]{Barr:1990td}%
  \BibitemOpen
  \bibfield  {author} {\bibinfo {author} {\bibfnamefont {S.~M.}\ \bibnamefont {Barr}},\ }\bibfield  {title} {\bibinfo {title} {{A Predictive hierarchical mode of quark and lepton masses}},\ }\href {https://doi.org/10.1103/PhysRevD.42.3150} {\bibfield  {journal} {\bibinfo  {journal} {Phys. Rev. D}\ }\textbf {\bibinfo {volume} {42}},\ \bibinfo {pages} {3150} (\bibinfo {year} {1990})}\BibitemShut {NoStop}%
\bibitem [{\citenamefont {Dobrescu}\ and\ \citenamefont {Fox}(2008)}]{Dobrescu:2008sz}%
  \BibitemOpen
  \bibfield  {author} {\bibinfo {author} {\bibfnamefont {B.~A.}\ \bibnamefont {Dobrescu}}\ and\ \bibinfo {author} {\bibfnamefont {P.~J.}\ \bibnamefont {Fox}},\ }\bibfield  {title} {\bibinfo {title} {{Quark and lepton masses from top loops}},\ }\href {https://doi.org/10.1088/1126-6708/2008/08/100} {\bibfield  {journal} {\bibinfo  {journal} {JHEP}\ }\textbf {\bibinfo {volume} {08}},\ \bibinfo {pages} {100}},\ \Eprint {https://arxiv.org/abs/0805.0822} {arXiv:0805.0822 [hep-ph]} \BibitemShut {NoStop}%
\bibitem [{\citenamefont {Babu}\ and\ \citenamefont {Thapa}(2020)}]{Babu:2020bgz}%
  \BibitemOpen
  \bibfield  {author} {\bibinfo {author} {\bibfnamefont {K.~S.}\ \bibnamefont {Babu}}\ and\ \bibinfo {author} {\bibfnamefont {A.}~\bibnamefont {Thapa}},\ }\bibfield  {title} {\bibinfo {title} {{Left-Right Symmetric Model without Higgs Triplets}},\ }\href@noop {} {\  (\bibinfo {year} {2020})},\ \Eprint {https://arxiv.org/abs/2012.13420} {arXiv:2012.13420 [hep-ph]} \BibitemShut {NoStop}%
\bibitem [{\citenamefont {Mohanta}\ and\ \citenamefont {Patel}(2022)}]{PhysRevD.106.075020}%
  \BibitemOpen
  \bibfield  {author} {\bibinfo {author} {\bibfnamefont {G.}~\bibnamefont {Mohanta}}\ and\ \bibinfo {author} {\bibfnamefont {K.~M.}\ \bibnamefont {Patel}},\ }\bibfield  {title} {\bibinfo {title} {Radiatively generated fermion mass hierarchy from flavor nonuniversal gauge symmetries},\ }\href {https://doi.org/10.1103/PhysRevD.106.075020} {\bibfield  {journal} {\bibinfo  {journal} {Phys. Rev. D}\ }\textbf {\bibinfo {volume} {106}},\ \bibinfo {pages} {075020} (\bibinfo {year} {2022})}\BibitemShut {NoStop}%
\bibitem [{\citenamefont {Kuchimanchi}\ and\ \citenamefont {Mohapatra}(1993)}]{PhysRevD.48.4352}%
  \BibitemOpen
  \bibfield  {author} {\bibinfo {author} {\bibfnamefont {R.}~\bibnamefont {Kuchimanchi}}\ and\ \bibinfo {author} {\bibfnamefont {R.~N.}\ \bibnamefont {Mohapatra}},\ }\bibfield  {title} {\bibinfo {title} {No parity violation without $r$-parity violation},\ }\href {https://doi.org/10.1103/PhysRevD.48.4352} {\bibfield  {journal} {\bibinfo  {journal} {Phys. Rev. D}\ }\textbf {\bibinfo {volume} {48}},\ \bibinfo {pages} {4352} (\bibinfo {year} {1993})}\BibitemShut {NoStop}%
\bibitem [{\citenamefont {Akhmedov}\ \emph {et~al.}(2024)\citenamefont {Akhmedov}, \citenamefont {{Bhupal Dev}}, \citenamefont {Jana},\ and\ \citenamefont {Mohapatra}}]{AKHMEDOV2024138616}%
  \BibitemOpen
  \bibfield  {author} {\bibinfo {author} {\bibfnamefont {E.}~\bibnamefont {Akhmedov}}, \bibinfo {author} {\bibfnamefont {P.}~\bibnamefont {{Bhupal Dev}}}, \bibinfo {author} {\bibfnamefont {S.}~\bibnamefont {Jana}},\ and\ \bibinfo {author} {\bibfnamefont {R.~N.}\ \bibnamefont {Mohapatra}},\ }\bibfield  {title} {\bibinfo {title} {Long-lived doubly charged scalars in the left-right symmetric model: Catalyzed nuclear fusion and collider implications},\ }\href {https://doi.org/https://doi.org/10.1016/j.physletb.2024.138616} {\bibfield  {journal} {\bibinfo  {journal} {Physics Letters B}\ }\textbf {\bibinfo {volume} {852}},\ \bibinfo {pages} {138616} (\bibinfo {year} {2024})}\BibitemShut {NoStop}%
\bibitem [{\citenamefont {Chakrabortty}\ \emph {et~al.}(2016)\citenamefont {Chakrabortty}, \citenamefont {Gluza}, \citenamefont {Jeliński},\ and\ \citenamefont {Srivastava}}]{CHAKRABORTTY2016361}%
  \BibitemOpen
  \bibfield  {author} {\bibinfo {author} {\bibfnamefont {J.}~\bibnamefont {Chakrabortty}}, \bibinfo {author} {\bibfnamefont {J.}~\bibnamefont {Gluza}}, \bibinfo {author} {\bibfnamefont {T.}~\bibnamefont {Jeliński}},\ and\ \bibinfo {author} {\bibfnamefont {T.}~\bibnamefont {Srivastava}},\ }\bibfield  {title} {\bibinfo {title} {Theoretical constraints on masses of heavy particles in left-right symmetric models},\ }\href {https://doi.org/https://doi.org/10.1016/j.physletb.2016.05.092} {\bibfield  {journal} {\bibinfo  {journal} {Physics Letters B}\ }\textbf {\bibinfo {volume} {759}},\ \bibinfo {pages} {361} (\bibinfo {year} {2016})}\BibitemShut {NoStop}%
\bibitem [{\citenamefont {Dev}\ \emph {et~al.}(2016)\citenamefont {Dev}, \citenamefont {Mohapatra},\ and\ \citenamefont {Zhang}}]{Dev2016}%
  \BibitemOpen
  \bibfield  {author} {\bibinfo {author} {\bibfnamefont {P.~S.~B.}\ \bibnamefont {Dev}}, \bibinfo {author} {\bibfnamefont {R.~N.}\ \bibnamefont {Mohapatra}},\ and\ \bibinfo {author} {\bibfnamefont {Y.}~\bibnamefont {Zhang}},\ }\bibfield  {title} {\bibinfo {title} {Probing the higgs sector of the minimal left-right symmetric model at future hadron colliders},\ }\href {https://doi.org/10.1007/JHEP05(2016)174} {\bibfield  {journal} {\bibinfo  {journal} {Journal of High Energy Physics}\ }\textbf {\bibinfo {volume} {2016}},\ \bibinfo {pages} {174} (\bibinfo {year} {2016})}\BibitemShut {NoStop}%
\bibitem [{\citenamefont {Chang}\ \emph {et~al.}(1984)\citenamefont {Chang}, \citenamefont {Mohapatra},\ and\ \citenamefont {Parida}}]{PhysRevLett.52.1072}%
  \BibitemOpen
  \bibfield  {author} {\bibinfo {author} {\bibfnamefont {D.}~\bibnamefont {Chang}}, \bibinfo {author} {\bibfnamefont {R.~N.}\ \bibnamefont {Mohapatra}},\ and\ \bibinfo {author} {\bibfnamefont {M.~K.}\ \bibnamefont {Parida}},\ }\bibfield  {title} {\bibinfo {title} {Decoupling of parity- and $\mathrm{SU}{(2)}_{R}$-breaking scales: A new approach to left-right symmetric models},\ }\href {https://doi.org/10.1103/PhysRevLett.52.1072} {\bibfield  {journal} {\bibinfo  {journal} {Phys. Rev. Lett.}\ }\textbf {\bibinfo {volume} {52}},\ \bibinfo {pages} {1072} (\bibinfo {year} {1984})}\BibitemShut {NoStop}%
\bibitem [{\citenamefont {Kuchimanchi}(2015)}]{Kuchimanchi_2015}%
  \BibitemOpen
  \bibfield  {author} {\bibinfo {author} {\bibfnamefont {R.}~\bibnamefont {Kuchimanchi}},\ }\bibfield  {title} {\bibinfo {title} {{Leptonic CP problem in left-right symmetric model}},\ }\href {https://doi.org/10.1103/PhysRevD.91.071901} {\bibfield  {journal} {\bibinfo  {journal} {Phys. Rev. D}\ }\textbf {\bibinfo {volume} {91}},\ \bibinfo {pages} {071901(R)} (\bibinfo {year} {2015})},\ \Eprint {https://arxiv.org/abs/1408.6382} {arXiv:1408.6382 [hep-ph]} \BibitemShut {NoStop}%
\bibitem [{\citenamefont {Abel}\ \emph {et~al.}(2020)\citenamefont {Abel}, \citenamefont {Afach}, \citenamefont {Ayres}, \citenamefont {Baker}, \citenamefont {Ban}, \citenamefont {Bison}, \citenamefont {Bodek}, \citenamefont {Bondar}, \citenamefont {Burghoff}, \citenamefont {Chanel}, \citenamefont {Chowdhuri}, \citenamefont {Chiu}, \citenamefont {Clement}, \citenamefont {Crawford}, \citenamefont {Daum}, \citenamefont {Emmenegger}, \citenamefont {Ferraris-Bouchez}, \citenamefont {Fertl}, \citenamefont {Flaux}, \citenamefont {Franke}, \citenamefont {Fratangelo}, \citenamefont {Geltenbort}, \citenamefont {Green}, \citenamefont {Griffith}, \citenamefont {van~der Grinten}, \citenamefont {Gruji\ifmmode~\acute{c}\else \'{c}\fi{}}, \citenamefont {Harris}, \citenamefont {Hayen}, \citenamefont {Heil}, \citenamefont {Henneck}, \citenamefont {H\'elaine}, \citenamefont {Hild}, \citenamefont {Hodge}, \citenamefont {Horras}, \citenamefont {Iaydjiev}, \citenamefont {Ivanov}, \citenamefont {Kasprzak}, \citenamefont
  {Kermaidic}, \citenamefont {Kirch}, \citenamefont {Knecht}, \citenamefont {Knowles}, \citenamefont {Koch}, \citenamefont {Koss}, \citenamefont {Komposch}, \citenamefont {Kozela}, \citenamefont {Kraft}, \citenamefont {Krempel}, \citenamefont {Ku\ifmmode~\acute{z}\else \'{z}\fi{}niak}, \citenamefont {Lauss}, \citenamefont {Lefort}, \citenamefont {Lemi\`ere}, \citenamefont {Leredde}, \citenamefont {Mohanmurthy}, \citenamefont {Mtchedlishvili}, \citenamefont {Musgrave}, \citenamefont {Naviliat-Cuncic}, \citenamefont {Pais}, \citenamefont {Piegsa}, \citenamefont {Pierre}, \citenamefont {Pignol}, \citenamefont {Plonka-Spehr}, \citenamefont {Prashanth}, \citenamefont {Qu\'em\'ener}, \citenamefont {Rawlik}, \citenamefont {Rebreyend}, \citenamefont {Rien\"acker}, \citenamefont {Ries}, \citenamefont {Roccia}, \citenamefont {Rogel}, \citenamefont {Rozpedzik}, \citenamefont {Schnabel}, \citenamefont {Schmidt-Wellenburg}, \citenamefont {Severijns}, \citenamefont {Shiers}, \citenamefont {Tavakoli~Dinani}, \citenamefont
  {Thorne}, \citenamefont {Virot}, \citenamefont {Voigt}, \citenamefont {Weis}, \citenamefont {Wursten}, \citenamefont {Wyszynski}, \citenamefont {Zejma}, \citenamefont {Zenner},\ and\ \citenamefont {Zsigmond}}]{PhysRevLett.124.081803}%
  \BibitemOpen
  \bibfield  {author} {\bibinfo {author} {\bibfnamefont {C.}~\bibnamefont {Abel}}, \bibinfo {author} {\bibfnamefont {S.}~\bibnamefont {Afach}}, \bibinfo {author} {\bibfnamefont {N.~J.}\ \bibnamefont {Ayres}}, \bibinfo {author} {\bibfnamefont {C.~A.}\ \bibnamefont {Baker}}, \bibinfo {author} {\bibfnamefont {G.}~\bibnamefont {Ban}}, \bibinfo {author} {\bibfnamefont {G.}~\bibnamefont {Bison}}, \bibinfo {author} {\bibfnamefont {K.}~\bibnamefont {Bodek}}, \bibinfo {author} {\bibfnamefont {V.}~\bibnamefont {Bondar}}, \bibinfo {author} {\bibfnamefont {M.}~\bibnamefont {Burghoff}}, \bibinfo {author} {\bibfnamefont {E.}~\bibnamefont {Chanel}}, \bibinfo {author} {\bibfnamefont {Z.}~\bibnamefont {Chowdhuri}}, \bibinfo {author} {\bibfnamefont {P.-J.}\ \bibnamefont {Chiu}}, \bibinfo {author} {\bibfnamefont {B.}~\bibnamefont {Clement}}, \bibinfo {author} {\bibfnamefont {C.~B.}\ \bibnamefont {Crawford}}, \bibinfo {author} {\bibfnamefont {M.}~\bibnamefont {Daum}}, \bibinfo {author} {\bibfnamefont {S.}~\bibnamefont
  {Emmenegger}}, \bibinfo {author} {\bibfnamefont {L.}~\bibnamefont {Ferraris-Bouchez}}, \bibinfo {author} {\bibfnamefont {M.}~\bibnamefont {Fertl}}, \bibinfo {author} {\bibfnamefont {P.}~\bibnamefont {Flaux}}, \bibinfo {author} {\bibfnamefont {B.}~\bibnamefont {Franke}}, \bibinfo {author} {\bibfnamefont {A.}~\bibnamefont {Fratangelo}}, \bibinfo {author} {\bibfnamefont {P.}~\bibnamefont {Geltenbort}}, \bibinfo {author} {\bibfnamefont {K.}~\bibnamefont {Green}}, \bibinfo {author} {\bibfnamefont {W.~C.}\ \bibnamefont {Griffith}}, \bibinfo {author} {\bibfnamefont {M.}~\bibnamefont {van~der Grinten}}, \bibinfo {author} {\bibfnamefont {Z.~D.}\ \bibnamefont {Gruji\ifmmode~\acute{c}\else \'{c}\fi{}}}, \bibinfo {author} {\bibfnamefont {P.~G.}\ \bibnamefont {Harris}}, \bibinfo {author} {\bibfnamefont {L.}~\bibnamefont {Hayen}}, \bibinfo {author} {\bibfnamefont {W.}~\bibnamefont {Heil}}, \bibinfo {author} {\bibfnamefont {R.}~\bibnamefont {Henneck}}, \bibinfo {author} {\bibfnamefont {V.}~\bibnamefont {H\'elaine}},
  \bibinfo {author} {\bibfnamefont {N.}~\bibnamefont {Hild}}, \bibinfo {author} {\bibfnamefont {Z.}~\bibnamefont {Hodge}}, \bibinfo {author} {\bibfnamefont {M.}~\bibnamefont {Horras}}, \bibinfo {author} {\bibfnamefont {P.}~\bibnamefont {Iaydjiev}}, \bibinfo {author} {\bibfnamefont {S.~N.}\ \bibnamefont {Ivanov}}, \bibinfo {author} {\bibfnamefont {M.}~\bibnamefont {Kasprzak}}, \bibinfo {author} {\bibfnamefont {Y.}~\bibnamefont {Kermaidic}}, \bibinfo {author} {\bibfnamefont {K.}~\bibnamefont {Kirch}}, \bibinfo {author} {\bibfnamefont {A.}~\bibnamefont {Knecht}}, \bibinfo {author} {\bibfnamefont {P.}~\bibnamefont {Knowles}}, \bibinfo {author} {\bibfnamefont {H.-C.}\ \bibnamefont {Koch}}, \bibinfo {author} {\bibfnamefont {P.~A.}\ \bibnamefont {Koss}}, \bibinfo {author} {\bibfnamefont {S.}~\bibnamefont {Komposch}}, \bibinfo {author} {\bibfnamefont {A.}~\bibnamefont {Kozela}}, \bibinfo {author} {\bibfnamefont {A.}~\bibnamefont {Kraft}}, \bibinfo {author} {\bibfnamefont {J.}~\bibnamefont {Krempel}}, \bibinfo
  {author} {\bibfnamefont {M.}~\bibnamefont {Ku\ifmmode~\acute{z}\else \'{z}\fi{}niak}}, \bibinfo {author} {\bibfnamefont {B.}~\bibnamefont {Lauss}}, \bibinfo {author} {\bibfnamefont {T.}~\bibnamefont {Lefort}}, \bibinfo {author} {\bibfnamefont {Y.}~\bibnamefont {Lemi\`ere}}, \bibinfo {author} {\bibfnamefont {A.}~\bibnamefont {Leredde}}, \bibinfo {author} {\bibfnamefont {P.}~\bibnamefont {Mohanmurthy}}, \bibinfo {author} {\bibfnamefont {A.}~\bibnamefont {Mtchedlishvili}}, \bibinfo {author} {\bibfnamefont {M.}~\bibnamefont {Musgrave}}, \bibinfo {author} {\bibfnamefont {O.}~\bibnamefont {Naviliat-Cuncic}}, \bibinfo {author} {\bibfnamefont {D.}~\bibnamefont {Pais}}, \bibinfo {author} {\bibfnamefont {F.~M.}\ \bibnamefont {Piegsa}}, \bibinfo {author} {\bibfnamefont {E.}~\bibnamefont {Pierre}}, \bibinfo {author} {\bibfnamefont {G.}~\bibnamefont {Pignol}}, \bibinfo {author} {\bibfnamefont {C.}~\bibnamefont {Plonka-Spehr}}, \bibinfo {author} {\bibfnamefont {P.~N.}\ \bibnamefont {Prashanth}}, \bibinfo {author}
  {\bibfnamefont {G.}~\bibnamefont {Qu\'em\'ener}}, \bibinfo {author} {\bibfnamefont {M.}~\bibnamefont {Rawlik}}, \bibinfo {author} {\bibfnamefont {D.}~\bibnamefont {Rebreyend}}, \bibinfo {author} {\bibfnamefont {I.}~\bibnamefont {Rien\"acker}}, \bibinfo {author} {\bibfnamefont {D.}~\bibnamefont {Ries}}, \bibinfo {author} {\bibfnamefont {S.}~\bibnamefont {Roccia}}, \bibinfo {author} {\bibfnamefont {G.}~\bibnamefont {Rogel}}, \bibinfo {author} {\bibfnamefont {D.}~\bibnamefont {Rozpedzik}}, \bibinfo {author} {\bibfnamefont {A.}~\bibnamefont {Schnabel}}, \bibinfo {author} {\bibfnamefont {P.}~\bibnamefont {Schmidt-Wellenburg}}, \bibinfo {author} {\bibfnamefont {N.}~\bibnamefont {Severijns}}, \bibinfo {author} {\bibfnamefont {D.}~\bibnamefont {Shiers}}, \bibinfo {author} {\bibfnamefont {R.}~\bibnamefont {Tavakoli~Dinani}}, \bibinfo {author} {\bibfnamefont {J.~A.}\ \bibnamefont {Thorne}}, \bibinfo {author} {\bibfnamefont {R.}~\bibnamefont {Virot}}, \bibinfo {author} {\bibfnamefont {J.}~\bibnamefont {Voigt}},
  \bibinfo {author} {\bibfnamefont {A.}~\bibnamefont {Weis}}, \bibinfo {author} {\bibfnamefont {E.}~\bibnamefont {Wursten}}, \bibinfo {author} {\bibfnamefont {G.}~\bibnamefont {Wyszynski}}, \bibinfo {author} {\bibfnamefont {J.}~\bibnamefont {Zejma}}, \bibinfo {author} {\bibfnamefont {J.}~\bibnamefont {Zenner}},\ and\ \bibinfo {author} {\bibfnamefont {G.}~\bibnamefont {Zsigmond}},\ }\bibfield  {title} {\bibinfo {title} {Measurement of the permanent electric dipole moment of the neutron},\ }\href {https://doi.org/10.1103/PhysRevLett.124.081803} {\bibfield  {journal} {\bibinfo  {journal} {Phys. Rev. Lett.}\ }\textbf {\bibinfo {volume} {124}},\ \bibinfo {pages} {081803} (\bibinfo {year} {2020})}\BibitemShut {NoStop}%
\bibitem [{\citenamefont {Abe}\ and\ \citenamefont {Tanaka}(2024)}]{10.3389/fphy.2024.1378254}%
  \BibitemOpen
  \bibfield  {author} {\bibinfo {author} {\bibfnamefont {K.}~\bibnamefont {Abe}}\ and\ \bibinfo {author} {\bibfnamefont {H.-K.}\ \bibnamefont {Tanaka}},\ }\bibfield  {title} {\bibinfo {title} {Hyper-kamiokande construction status and prospects},\ }\bibfield  {journal} {\bibinfo  {journal} {Frontiers in Physics}\ }\textbf {\bibinfo {volume} {12}},\ \href {https://doi.org/10.3389/fphy.2024.1378254} {10.3389/fphy.2024.1378254} (\bibinfo {year} {2024})\BibitemShut {NoStop}%
\bibitem [{\citenamefont {Terranova}(2024)}]{universe10050221}%
  \BibitemOpen
  \bibfield  {author} {\bibinfo {author} {\bibfnamefont {F.}~\bibnamefont {Terranova}},\ }\bibfield  {title} {\bibinfo {title} {Future long-baseline neutrino experiments},\ }\bibfield  {journal} {\bibinfo  {journal} {Universe}\ }\textbf {\bibinfo {volume} {10}},\ \href {https://doi.org/10.3390/universe10050221} {10.3390/universe10050221} (\bibinfo {year} {2024})\BibitemShut {NoStop}%
\bibitem [{\citenamefont {Vallari}(2024)}]{zoya}%
  \BibitemOpen
  \bibfield  {author} {\bibinfo {author} {\bibfnamefont {Z.}~\bibnamefont {Vallari}},\ }\bibfield  {title} {\bibinfo {title} {{Results from a joint analysis of data from NOvA and T2K}},\ }\href@noop {} {\  (\bibinfo {year} {2024})},\ \Eprint {https://arxiv.org/abs/Seminar: https://indico.fnal.gov/event/62062/} {Seminar: https://indico.fnal.gov/event/62062/} \BibitemShut {NoStop}%
\bibitem [{\citenamefont {Esteban}\ \emph {et~al.}(2020)\citenamefont {Esteban}, \citenamefont {Gonzalez-Garcia}, \citenamefont {Maltoni}, \citenamefont {Schwetz},\ and\ \citenamefont {Zhou}}]{Esteban_2020}%
  \BibitemOpen
  \bibfield  {author} {\bibinfo {author} {\bibfnamefont {I.}~\bibnamefont {Esteban}}, \bibinfo {author} {\bibfnamefont {M.}~\bibnamefont {Gonzalez-Garcia}}, \bibinfo {author} {\bibfnamefont {M.}~\bibnamefont {Maltoni}}, \bibinfo {author} {\bibfnamefont {T.}~\bibnamefont {Schwetz}},\ and\ \bibinfo {author} {\bibfnamefont {A.}~\bibnamefont {Zhou}},\ }\bibfield  {title} {\bibinfo {title} {The fate of hints: updated global analysis of three-flavor neutrino oscillations},\ }\bibfield  {journal} {\bibinfo  {journal} {Journal of High Energy Physics}\ }\textbf {\bibinfo {volume} {2020}},\ \href {https://doi.org/10.1007/jhep09(2020)178} {10.1007/jhep09(2020)178} (\bibinfo {year} {2020})\BibitemShut {NoStop}%
\bibitem [{\citenamefont {Esteban}\ \emph {et~al.}(2024)\citenamefont {Esteban}, \citenamefont {Gonzalez-Garcia}, \citenamefont {Maltoni}, \citenamefont {Martinez-Soler}, \citenamefont {Pinheiro},\ and\ \citenamefont {Schwetz}}]{esteban2024nufit60updatedglobalanalysis}%
  \BibitemOpen
  \bibfield  {author} {\bibinfo {author} {\bibfnamefont {I.}~\bibnamefont {Esteban}}, \bibinfo {author} {\bibfnamefont {M.~C.}\ \bibnamefont {Gonzalez-Garcia}}, \bibinfo {author} {\bibfnamefont {M.}~\bibnamefont {Maltoni}}, \bibinfo {author} {\bibfnamefont {I.}~\bibnamefont {Martinez-Soler}}, \bibinfo {author} {\bibfnamefont {J.~P.}\ \bibnamefont {Pinheiro}},\ and\ \bibinfo {author} {\bibfnamefont {T.}~\bibnamefont {Schwetz}},\ }\href {https://arxiv.org/abs/2410.05380} {\bibinfo {title} {Nufit-6.0: Updated global analysis of three-flavor neutrino oscillations}} (\bibinfo {year} {2024}),\ \Eprint {https://arxiv.org/abs/2410.05380} {arXiv:2410.05380 [hep-ph]} \BibitemShut {NoStop}%
\bibitem [{\citenamefont {Kuchimanchi}(2010)}]{Kuchimanchi:2010xs}%
  \BibitemOpen
  \bibfield  {author} {\bibinfo {author} {\bibfnamefont {R.}~\bibnamefont {Kuchimanchi}},\ }\bibfield  {title} {\bibinfo {title} {{$P/CP$ Conserving $CP/P$ Violation Solves Strong $CP$ Problem}},\ }\href {https://doi.org/10.1103/PhysRevD.82.116008} {\bibfield  {journal} {\bibinfo  {journal} {Phys. Rev.}\ }\textbf {\bibinfo {volume} {D82}},\ \bibinfo {pages} {116008} (\bibinfo {year} {2010})},\ \Eprint {https://arxiv.org/abs/1009.5961} {arXiv:1009.5961 [hep-ph]} \BibitemShut {NoStop}%
%%CITATION = 1009.5961;%%
\bibitem [{\citenamefont {Kuchimanchi}(2023)}]{PhysRevD.108.095023}%
  \BibitemOpen
  \bibfield  {author} {\bibinfo {author} {\bibfnamefont {R.}~\bibnamefont {Kuchimanchi}},\ }\bibfield  {title} {\bibinfo {title} {$p$ and $cp$ solution of the strong $cp$ puzzle},\ }\href {https://doi.org/10.1103/PhysRevD.108.095023} {\bibfield  {journal} {\bibinfo  {journal} {Phys. Rev. D}\ }\textbf {\bibinfo {volume} {108}},\ \bibinfo {pages} {095023} (\bibinfo {year} {2023})}\BibitemShut {NoStop}%
\bibitem [{\citenamefont {Kuchimanchi}(2014)}]{Kuchimanchi:2012te}%
  \BibitemOpen
  \bibfield  {author} {\bibinfo {author} {\bibfnamefont {R.}~\bibnamefont {Kuchimanchi}},\ }\bibfield  {title} {\bibinfo {title} {P stabilizes dark matter and with cp can predict leptonic phases},\ }\href {https://doi.org/10.1140/epjc/s10052-014-2726-5} {\bibfield  {journal} {\bibinfo  {journal} {Eur. Phys. J. C}\ }\textbf {\bibinfo {volume} {74}},\ \bibinfo {pages} {1} (\bibinfo {year} {2014})}\BibitemShut {NoStop}%
\bibitem [{\citenamefont {{Peccei}}\ and\ \citenamefont {{Quinn}}(1977)}]{1977PhRvL..38.1440P}%
  \BibitemOpen
  \bibfield  {author} {\bibinfo {author} {\bibfnamefont {R.~D.}\ \bibnamefont {{Peccei}}}\ and\ \bibinfo {author} {\bibfnamefont {H.~R.}\ \bibnamefont {{Quinn}}},\ }\bibfield  {title} {\bibinfo {title} {{CP conservation in the presence of pseudoparticles}},\ }\href {https://doi.org/10.1103/PhysRevLett.38.1440} {\bibfield  {journal} {\bibinfo  {journal} {\prl}\ }\textbf {\bibinfo {volume} {38}},\ \bibinfo {pages} {1440} (\bibinfo {year} {1977})}\BibitemShut {NoStop}%
\bibitem [{\citenamefont {Weinberg}(1978)}]{PhysRevLett.40.223}%
  \BibitemOpen
  \bibfield  {author} {\bibinfo {author} {\bibfnamefont {S.}~\bibnamefont {Weinberg}},\ }\bibfield  {title} {\bibinfo {title} {A new light boson?},\ }\href {https://doi.org/10.1103/PhysRevLett.40.223} {\bibfield  {journal} {\bibinfo  {journal} {Phys. Rev. Lett.}\ }\textbf {\bibinfo {volume} {40}},\ \bibinfo {pages} {223} (\bibinfo {year} {1978})}\BibitemShut {NoStop}%
\bibitem [{\citenamefont {Wilczek}(1978)}]{PhysRevLett.40.279}%
  \BibitemOpen
  \bibfield  {author} {\bibinfo {author} {\bibfnamefont {F.}~\bibnamefont {Wilczek}},\ }\bibfield  {title} {\bibinfo {title} {Problem of strong $p$ and $t$ invariance in the presence of instantons},\ }\href {https://doi.org/10.1103/PhysRevLett.40.279} {\bibfield  {journal} {\bibinfo  {journal} {Phys. Rev. Lett.}\ }\textbf {\bibinfo {volume} {40}},\ \bibinfo {pages} {279} (\bibinfo {year} {1978})}\BibitemShut {NoStop}%
\bibitem [{\citenamefont {Maiezza}\ and\ \citenamefont {Nesti}(2022)}]{Maiezza:2021dui}%
  \BibitemOpen
  \bibfield  {author} {\bibinfo {author} {\bibfnamefont {A.}~\bibnamefont {Maiezza}}\ and\ \bibinfo {author} {\bibfnamefont {F.}~\bibnamefont {Nesti}},\ }\bibfield  {title} {\bibinfo {title} {{Parity from gauge symmetry}},\ }\href {https://doi.org/10.1140/epjc/s10052-022-10390-1} {\bibfield  {journal} {\bibinfo  {journal} {Eur. Phys. J. C}\ }\textbf {\bibinfo {volume} {82}},\ \bibinfo {pages} {491} (\bibinfo {year} {2022})},\ \Eprint {https://arxiv.org/abs/2111.11076} {arXiv:2111.11076 [hep-th]} \BibitemShut {NoStop}%
\end{thebibliography}%
%\end{flushleft} 

%\bibliography{LRsymmetrythi}

\end{document}